\documentclass{article}

\usepackage{microtype}
\usepackage{graphicx}
\usepackage{subcaption}
\usepackage{booktabs} 

\usepackage{hyperref}

\usepackage[preprint]{icml2026}

\usepackage{amsmath}
\usepackage{amssymb}
\usepackage{mathtools}
\usepackage{amsthm}

\usepackage[capitalize,noabbrev]{cleveref}

\usepackage{graphicx}
\usepackage{caption}
\usepackage{booktabs}
\usepackage{array}
\usepackage{float}
\usepackage{multicol}
\usepackage{multirow}
\usepackage{makecell}
\usepackage{colortbl}
\usepackage{pifont}
\usepackage{xcolor}
\usepackage{capt-of}
\usepackage{subcaption}
\usepackage{longtable}

\definecolor{blue}{RGB}{189,211,238} 
\definecolor{green}{RGB}{223,234,246}
\definecolor{yellow}{RGB}{200,222,218} 
\definecolor{red}{RGB}{252,248,225}

\definecolor{brightred}{RGB}{230,0,0}
\definecolor{brightgreen}{RGB}{0,230,0}

\usepackage{xurl}

\theoremstyle{plain}

\theoremstyle{definition}

\theoremstyle{remark}

\usepackage[textsize=tiny]{todonotes}

\icmltitlerunning{ProSoftArena: Benchmarking Hierarchical Capabilities of Multimodal Agents in Professional Software Environments}

\begin{document}

\twocolumn[
  \icmltitle{ProSoftArena: Benchmarking Hierarchical Capabilities of Multimodal \\Agents in Professional Software Environments}
    


  \icmlsetsymbol{equal}{*}

  \vspace{-3mm}
  \begin{icmlauthorlist}
    \icmlauthor{Jiaxin Ai}{whu,sii}
    \icmlauthor{Yukang Feng}{nku,sii}
    \icmlauthor{Fanrui Zhang}{ustc,sii}
    \icmlauthor{Jianwen Sun}{nku,sii}
    \icmlauthor{Zizhen Li}{nku,sii}
    \icmlauthor{Chuanhao Li}{pjlab}
    \icmlauthor{Yifan Chang}{ustc,sii}
    \icmlauthor{Wenxiao Wu}{hust,sii}
    \icmlauthor{Ruoxi Wang}{piz}
    \icmlauthor{Mingliang Zhai}{ustb}
    \icmlauthor{Kaipeng Zhang}{pjlab,sii}
  \end{icmlauthorlist}
    
    \icmlaffiliation{whu}{WHU}
    \icmlaffiliation{sii}{Shanghai Innovation Institute}
    \icmlaffiliation{nku}{NKU}
    \icmlaffiliation{ustc}{USTC}
    \icmlaffiliation{pjlab}{Shanghai AI Lab}
    \icmlaffiliation{piz}{PITT}
    \icmlaffiliation{hust}{HUST}
    \icmlaffiliation{ustb}{USTB}
    
    
    \icmlcorrespondingauthor{Kaipeng Zhang}{zhangkaipeng@pjlab.org.cn}

  \icmlkeywords{Computer-use Agents, MLLMs, Benchmarks, ICML}

  \vskip 0.15in
  
  \vbox{
    \centering
    \centerline{\includegraphics[width=\textwidth]{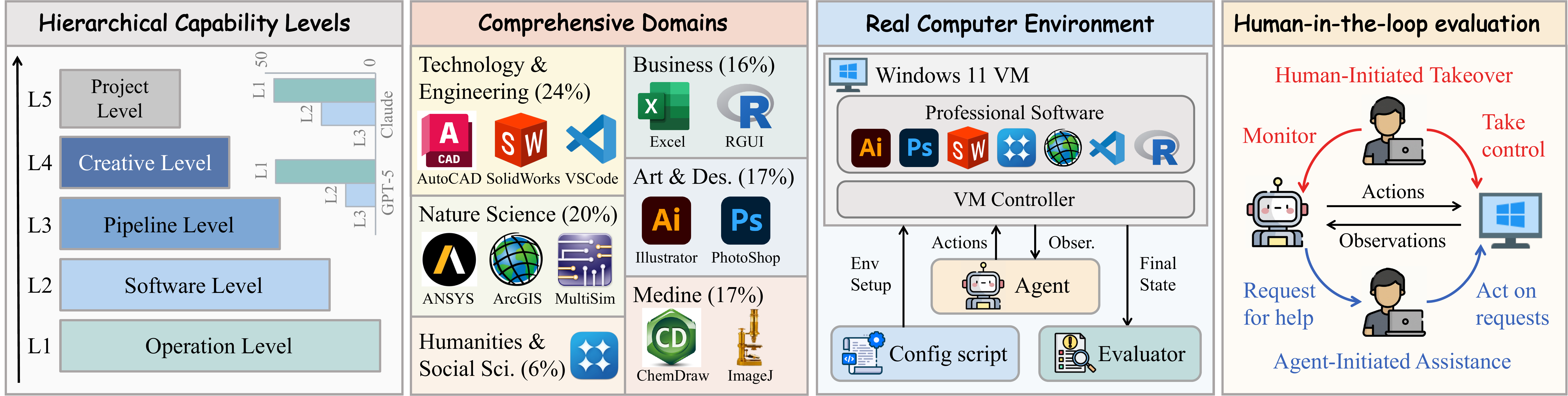}}
    \captionof{figure}{We establish the first hierarchical taxonomy of agent capabilities in professional software environments; and curate a comprehensive benchmark covering 6 disciplines, 20 subfields and 13 core professional applications. We construct a VM-based real computer environment for reproducible evaluations, and uniquely incorporate a human-in-the-loop evaluation paradigm.}
    \label{fig:overview}
  }
  \vskip 0.15in
]



\printAffiliationsAndNotice{}  

\begin{abstract}
Multimodal agents are making rapid progress on general computer-use tasks, yet existing benchmarks remain largely confined to browsers
and basic desktop applications, falling short in professional software workflows that dominate real-world scientific and industrial practice. 
To close this gap, we introduce ProSoftArena, a benchmark and platform specifically for evaluating multimodal agents in professional software environments.
We establish the first capability hierarchy tailored to agent use of professional software and construct a benchmark of 436 realistic work and research tasks spanning 6 disciplines and 13 core professional applications. 
To ensure reliable and reproducible assessment, we build an executable real-computer environment with an execution-based evaluation framework and uniquely incorporate a human-in-the-loop evaluation paradigm.
Extensive experiments show that even the best-performing agent attains only a 24.4\% success rate on L2 tasks and completely fails on L3 multi-software workflow. 
In-depth analysis further provides valuable insights for addressing current agent limitations and more effective design principles, paving the way to build more capable agents in professional software settings.
This project is available at: \small \url{https://prosoftarena.github.io}.
\end{abstract}


\section{Introduction}
\label{sec:intro}

Rapid advances in large vision-language models (VLMs) are catalyzing a paradigm shift toward generalist AI agents capable of perceiving, planning, and acting within digital environments~\cite{13_liu2018reinforcementlearningwebinterfaces,42_Schüpbach_San_Miguel_Ferchow_Meboldt_2025,43_wang2025opencuaopenfoundationscomputeruse}. By automating complex workflows through natural language instructions, such multimodal agents hold great potential to revolutionize human-computer interaction and significantly enhance productivity and accessibility~\cite{29_10.1007/978-3-031-73039-9_14,44_10.1007/978-3-031-73113-6_10,32_zhang-etal-2024-android,33_10.1145/3637528.3671620}. Accordingly, substantial research efforts have been devoted to developing and benchmarking such agents, with encouraging progress demonstrated on routine computer tasks—such as web navigation, and file management—using general-purpose software~\cite{22_sun2025surveyneuralcodeintelligence,44_10.1007/978-3-031-73113-6_10,45_xu-etal-2025-androidlab,46_song-etal-2025-beyond}

\begin{figure*}[t]
  \centering
  \includegraphics[width=\textwidth]{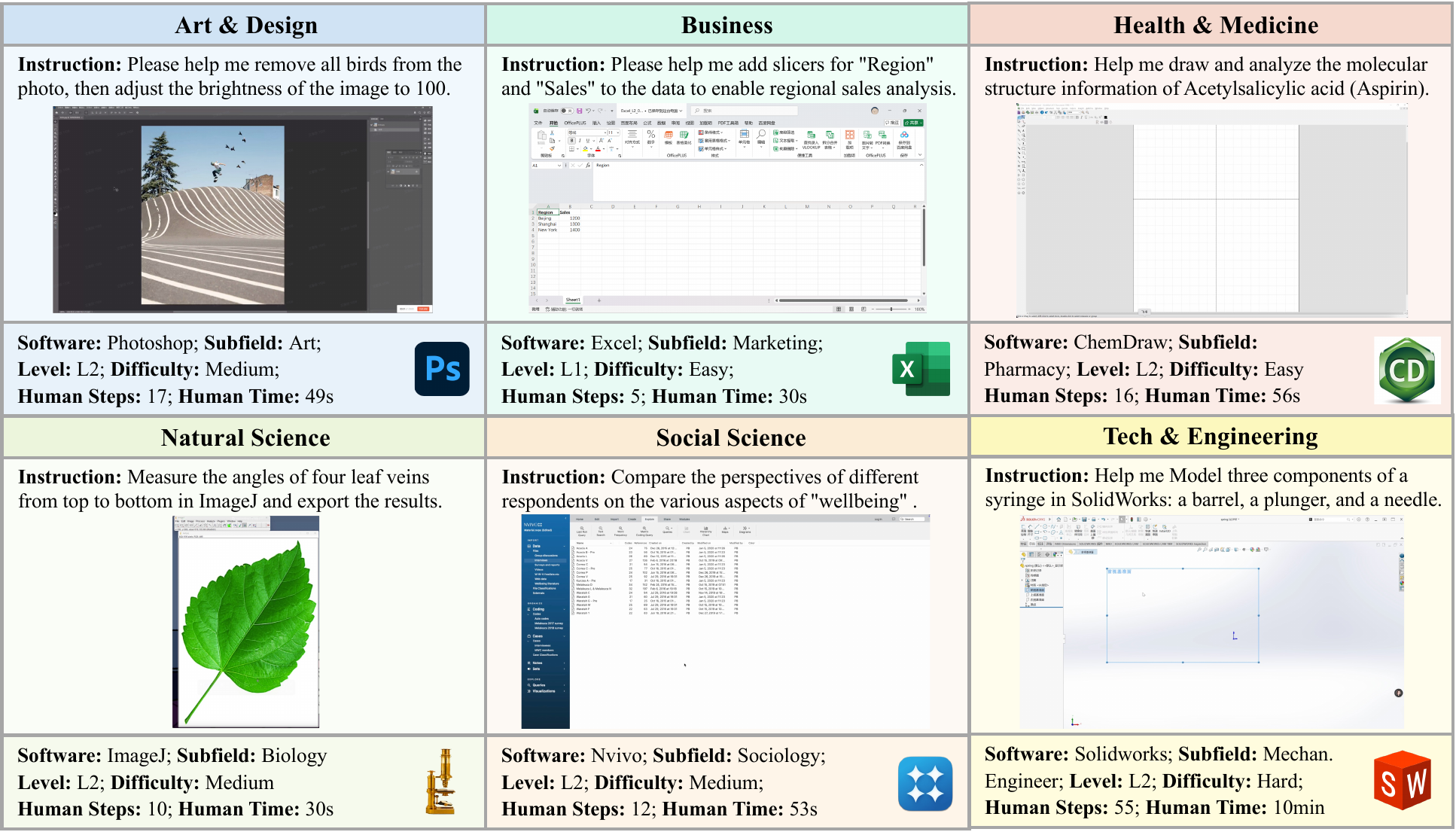}
  \caption{Representative Task Samples across Six Core Domains in ProSoftArena.
  For each domain, we showcase a typical task scenario including the specific natural language instruction, the initial computer state, and associated metadata.
  }
  \label{fig:domain_samples}
\end{figure*}

However, a critical yet largely unaddressed frontier remains: the mastery of professional software.~\cite{31_qin2025uitarspioneeringautomatedgui,47_xue2025illusionprogressassessingcurrent,48_davydova2025osuniversebenchmarkmultimodalguinavigation} In real-world scientific and industrial practice, experts 
rely on 
domain-specific software to execute core workflows~\cite{39_xie2025scalingcomputerusegroundinguser,50_u2024periguruperipheralroboticmobile,51_dai2025advancingmobileguiagents,41_yan2025mcpworldunifiedbenchmarkingtestbed,34_zhang2025ufo2desktopagentos}, such as SolidWorks for engineering design, ChemDraw for molecular modeling, and Adobe Suite for digital creation. These tools present a quantum leap in complexity compared to common desktop applications, featuring intricate graphical user interfaces (GUIs) and necessitating deep domain-expert knowledge for effective operation. Existing benchmarks, predominantly focused on daily tasks with general-purpose applications, fail to capture these challenges~\cite{7_deng-etal-2024-mobile,8_ICLR2025_01a83bc2,40_abhyankar2025osworldhumanbenchmarkingefficiencycomputeruse,42_Schüpbach_San_Miguel_Ferchow_Meboldt_2025,49_tian-etal-2025-mmina}. 
To bridge this gap, we introduce ProSoftArena, a comprehensive benchmark suite designed to systematically evaluate multimodal agents in professional software environments.


We establish a hierarchical taxonomy for agent capability in professional software: from basic GUI manipulation (L1) and software-level feature usage (L2) within a single application, to cross-application workflow execution (L3), and ultimately open-ended creation (L4) and real-world project-level orchestration (L5). This hierarchical categorization/structured framework not only enables systematic probing of the capability frontier in professional tools, but provides a well-defined roadmap for future research.

Building upon this taxonomy, we curate a benchmark with 436 tasks spanning 6 disciplines, 20 subfields, and 13 core professional applications. Figure~\ref{fig:domain_samples} presents
representative task samples from each domain.
Each task is meticulously designed by domain experts to simulate authentic professional workflows, requiring not only precise GUI interactions but also the application of deep, domain-specific knowledge for successful completion. Explicitly mapped to our five capability levels, the benchmark ensures comprehensive coverage from basic operations to complex project orchestration. Human performance studies confirm these tasks are more time-consuming and challenging than existing benchmarks, providing a rigorous measure of true capability in professional software environments.

\newcommand{\tabincell}[2]{\begin{tabular}{@{}#1@{}}#2\end{tabular}}

\begin{table*}[h]
  \centering
  \caption{Comparison between related benchmarks with our ProSoftArena.}
   \vspace{-1mm}
  \resizebox{\textwidth}{!}{
    \begin{tabular}{lccccccc}
        \toprule
         & \tabincell{c}{\textbf{\# Task} \\ \textbf{Samples}} 
         & \tabincell{c}{\textbf{Scenario} \\ \textbf{Coverage}} 
         & \tabincell{c}{\textbf{Hierarchical} \\ \textbf{Evaluation}} 
         & \tabincell{c}{\textbf{Self-Hosted} \\ \textbf{Environment}} 
         & \tabincell{c}{\textbf{Multi-} \\ \textbf{Discipline}} 
         & \tabincell{c}{\textbf{Human-in-loop} \\ \textbf{Evaluation}} 
         & \tabincell{c}{\textbf{\# Professional} \\ \textbf{Applications}} \\
        
        \midrule
        GAIA~\cite{gaia} & 466 & Daily & \textcolor{brightred}{\ding{55}} & \textcolor{brightred}{\ding{55}} & \textcolor{brightred}{\ding{55}} & \textcolor{brightred}{\ding{55}} & 0 \\
        OSWorld~\cite{11_OSWorld} & 369 & Daily & \textcolor{brightred}{\ding{55}} & \textcolor{brightgreen}{\ding{51}} & \textcolor{brightred}{\ding{55}} & \textcolor{brightred}{\ding{55}} & 3 \\
        WindowsAgentArena~\cite{12_bonatti2024windowsagentarenaevaluating} & 154 & Daily & \textcolor{brightred}{\ding{55}} & \textcolor{brightgreen}{\ding{51}} & \textcolor{brightred}{\ding{55}} & \textcolor{brightred}{\ding{55}} & 3 \\
        TheAgentCompany~\cite{18_xu2025theagentcompanybenchmarkingllmagents} & 175 & Work & \textcolor{brightred}{\ding{55}} & \textcolor{brightgreen}{\ding{51}} & \textcolor{brightred}{\ding{55}} & \textcolor{brightred}{\ding{55}} & 4 \\
        ScienceBoard~\cite{15_sun2025scienceboardevaluatingmultimodalautonomous} & 169 & Research & \textcolor{brightred}{\ding{55}} & \textcolor{brightgreen}{\ding{51}} & \textcolor{brightgreen}{\ding{51}} & \textcolor{brightred}{\ding{55}} & 6 \\
        \midrule
        \textsc{ProSoftArena} & 436 & Work \& Research & \textcolor{brightgreen}{\ding{51}} & \textcolor{brightgreen}{\ding{51}} & \textcolor{brightgreen}{\ding{51}} & \textcolor{brightgreen}{\ding{51}} & 13 \\ 
        \bottomrule
    \end{tabular}
  }
  \label{tab:comparison}
   \vspace{-3mm}
\end{table*}

To enable reliable and reproducible evaluation, we construct a unified platform that integrates an executable professional software environment with an automated assessment framework. The environment runs in isolated virtual machines mirroring real computer systems, pre-installed with required professional software. Agents perceive the environment through screen captures and system state signals, and interact via flexible keyboard/mouse control. The framework manages the entire evaluation lifecycle—from deterministic environment setup to iterative agent-environment interaction—culminating in execution-based evaluation via manually crafted scripts that automatically verify task completion by checking system internal states and output artifacts, thereby ensuring accurate and reliable assessment.


Our platform incorporates a human-in-the-loop evaluation paradigm. We implement two distinct collaboration modes: (i) Human-Initiated Takeover, where human experts can intervene at any point to correct a significant agent error or deviation; and (ii) Agent-Initiated Assistance, where agent proactively ask human for help when facing uncertainty. This enables a holistic evaluation of an agent's collaborative efficiency rather than just autonomous success.

We extensively evaluate state-of-the-art multimodal agents on ProSoftArena, including proprietary and open-source MLLMs as well as specialized computer-use agents. While agents achieve partial success on basic GUI operations (L1), performance drops sharply at software-level usage (L2), and cross-application workflows (L3) remain largely out of reach—highlighting critical gaps in long-horizon planning, state tracking, and semantic alignment across professional tools. Error analysis reveals consistent failure modes in task planning, domain knowledge, and visual grounding. Our ablations demonstrate that enriched visual inputs, domain priors, and longer action history can partially mitigate these issues, but also introduce non-trivial computational cost. Human-agent collaboration shows substantial practical value in professional software environments, yet current agents rarely request help under uncertainty, underscoring the need for calibrated self-assessment and proactive collaboration mechanisms in future systems.

\section{Related Work}
\label{sec:related}
\noindent
\textbf{Computer-using agents.} Computer-using agents aim to autonomously execute tasks in digital environments by interpreting natural language instructions and interacting with operating systems~\cite{11_OSWorld,12_bonatti2024windowsagentarenaevaluating,15_sun2025scienceboardevaluatingmultimodalautonomous,19_song2025mmaccopilotmultimodalagentcollaboration,20_zhang2025largelanguagemodelbrainedgui,36_hu-etal-2025-os}. Recent advances in multimodal large language models have driven progress along two main interaction paradigms: command-line–based agents that generate executable scripts~\cite{21_wu2024oscopilotgeneralistcomputeragents,22_sun2025surveyneuralcodeintelligence,25_wang2024executablecodeactionselicit,37_song2025coact1computerusingagentscoding}, and graphical user interface agents that perform human-like mouse and keyboard actions~\cite{11_OSWorld,23_zhang-etal-2025-ufo,24_10.1145/3706598.3713600,38_liu2025infiguir1advancingmultimodalgui}. Early systems like UFO established multi-agent architectures that combine vision–language models with accessibility APIs for cross-application coordination~\cite{19_song2025mmaccopilotmultimodalagentcollaboration,21_wu2024oscopilotgeneralistcomputeragents,23_zhang-etal-2025-ufo,26_zheng2024gpt4visiongeneralistwebagent} , while commercial systems including Claude Computer Use and OpenAI’s Operator demonstrate improved screenshot-based interaction through stronger multimodal reasoning ~\cite{35_hu2024dawnguiagentpreliminary,50_u2024periguruperipheralroboticmobile,51_dai2025advancingmobileguiagents}. Recent frameworks have further enhanced capabilities in planning ~\cite{6_koh-etal-2024-visualwebarena}, visual grounding ~\cite{1_Shi2017WorldOB,47_xue2025illusionprogressassessingcurrent,48_davydova2025osuniversebenchmarkmultimodalguinavigation}, and long-horizon control ~\cite{16_drouin2024workarenacapablewebagents,49_tian-etal-2025-mmina}. Despite this progress, existing research remains predominantly focused on general-purpose applications, leaving professional software agents largely unexplored. 
Our work firstly systematically investigates this high-value domain, 
constructing executable professional software environments and revealing critical limitations of current multimodal agents.

\noindent
\textbf{Benchmarks for Multimodal Agents.} Benchmarks for evaluating multimodal agents span web ~\cite{1_Shi2017WorldOB,2_liu2024visualwebbench,4_zhou2024webarenarealisticwebenvironment,5_deng2023mind2webgeneralistagentweb,6_koh-etal-2024-visualwebarena,49_tian-etal-2025-mmina}, mobile ~\cite{3_lu2025guiodysseycomprehensivedatasetcrossapp,7_deng-etal-2024-mobile,8_ICLR2025_01a83bc2,9_chai2025a3androidagentarena,50_u2024periguruperipheralroboticmobile,51_dai2025advancingmobileguiagents}, and desktop environments~\cite{10_NEURIPS2024_c2f71567,11_OSWorld,12_bonatti2024windowsagentarenaevaluating,44_10.1007/978-3-031-73113-6_10,48_davydova2025osuniversebenchmarkmultimodalguinavigation}. Early efforts focused on constrained short-horizon UI interactions, such as micro web tasks ~\cite{13_liu2018reinforcementlearningwebinterfaces} and smartphone operations ~\cite{8_ICLR2025_01a83bc2,9_chai2025a3androidagentarena,45_xu-etal-2025-androidlab}. Subsequent work introduced more realistic scenarios including open-ended web navigation~\cite{4_zhou2024webarenarealisticwebenvironment,6_koh-etal-2024-visualwebarena,46_song-etal-2025-beyond,47_xue2025illusionprogressassessingcurrent} and desktop task automation ~\cite{11_OSWorld,14_chen2025osmapfarcomputerusingagents,27_xu-etal-2025-crab,28_hong2024cogagentvisuallanguagemodel}. At the desktop level, OSWorld~\cite{11_OSWorld} and OS-MAP~\cite{14_chen2025osmapfarcomputerusingagents} cover multiple operating systems with diverse real-world tasks, while Windows Agent Arena~\cite{12_bonatti2024windowsagentarenaevaluating} provides scalable Windows-specific evaluation. Recent work also targets specialized contexts: ScienceBoard~\cite{15_sun2025scienceboardevaluatingmultimodalautonomous,30_yang-etal-2025-aria} evaluates scientific workflows, and WorkArena, WorkArena++, and TheAgentCompany~\cite{16_drouin2024workarenacapablewebagents,17_NEURIPS2024_0b82662b,18_xu2025theagentcompanybenchmarkingllmagents,47_xue2025illusionprogressassessingcurrent} assess knowledge-worker tasks with long-horizon projects. ProSoftArena first centers on professional software scenarios, establishing hierarchical capability levels for professional software use and constructing a testbed with 13 core applications across 20 disciplines. It further incorporates human-in-the-loop evaluation to assess agents’ collaborative performance.
\begin{figure*}[!t]
\centering
  \resizebox{1\linewidth}{!} {
    \includegraphics{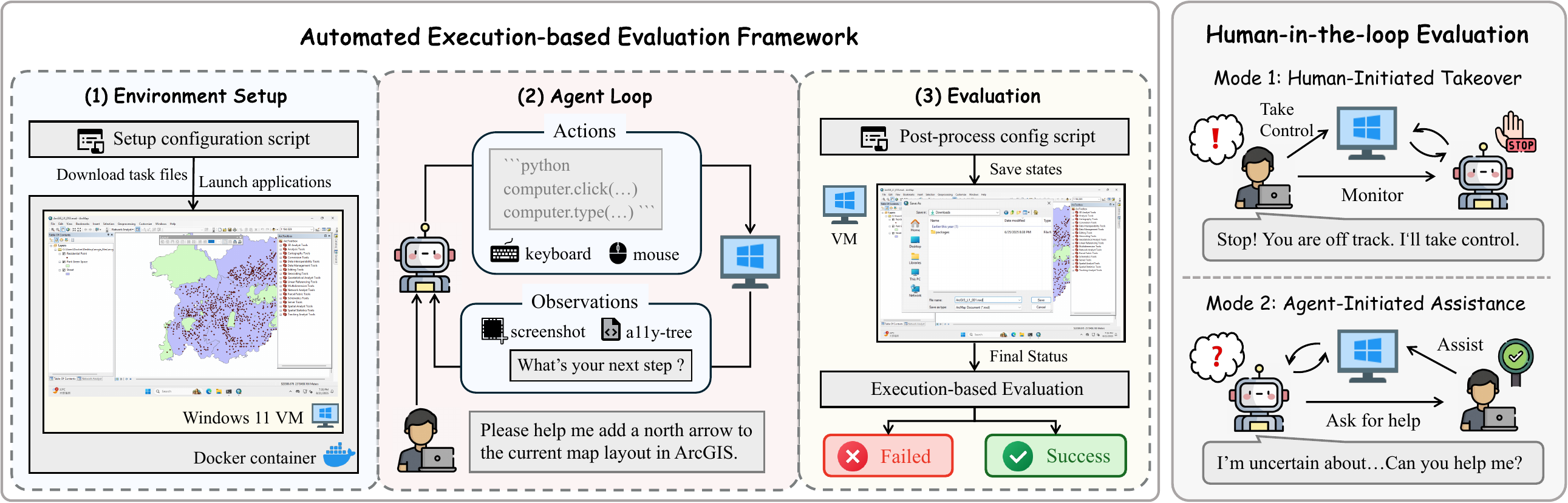}
  }
  \vspace{-5mm}
  \caption{Automated and Human-in-the-loop Evaluation Framework.
  }
    \label{fig:method}
  \vspace{-4mm}
\end{figure*}
\section{Hierarchical Capability Levels}

\label{sec:hierarchical}
We categorize the capabilities of agents operating professional software into hierarchical levels from L1 to L5, based on the complexity of decision-making and the depth of required domain expertise. Higher levels encompass the capabilities of lower tiers, defining a structured trajectory from atomic GUI operation to complex workflow execution. Our benchmark currently covers the assessment up to L4, leaving the exploration of project-level orchestration (L5) to future work. The detailed categorization of capability level is illustrated as below:

\noindent
\textbf{Operation Level} (L1): Agents at this capability level should possess the fundamental ability to understand the interface of professional software and perform basic operations proficiently. This entails executing atomic GUI interactions
to accomplish a single core function of the software, e.g., adjusting image brightness in Photoshop. This level forms the foundation for all higher-level skills.

\noindent
\textbf{Software Level} (L2): Agents at this capability level should be able to plan and execute a series of basic operations within the software to accomplish a complete professional task. This requires agents to make sequential decisions and apply domain knowledge. For example, draw a standard outdoor badminton court in AutoCAD. 

\noindent
\textbf{Pipeline Level} (L3): Progression to L3 entails the coordination of multiple software applications to execute cross-application workflows, delivering outcomes that meet specialized quality standards. This capability demands deep domain knowledge, long-horizon planning, and reliable context switching and data transfer across different applications. For example, aggregating street-level populations in ArcGIS and deriving distributional statistics in Excel for spatial demographic analysis.

\noindent
\textbf{Creative Level} (L4): Agents at this capability level should be able to leverage L2/L3 capabilities to accomplish open-ended creative tasks. For example, design a cartoon-style logo for a pediatric dentistry clinic that incorporates sun and smile elements. 

\noindent
\textbf{Project Level} (L5): At this capability level, agents (or agent systems) are expected to deliver end-to-end projects in real industrial and scientific settings. Such projects are typically long-horizon, collaborative and dynamic—spaning weeks or months, involving multi-department collaboration, and evolving as requirements change. This demands capabilities in long-horizon planning, multi-role coordination, dynamic adaptation to changing needs, and multi-stage workflow management across specialized domains. Achieving this level signifies the maturation of agents into fully autonomous digital workers for practical production and scientific research.

\begin{figure*}[t]
  \centering
  \includegraphics[width=0.95\textwidth]{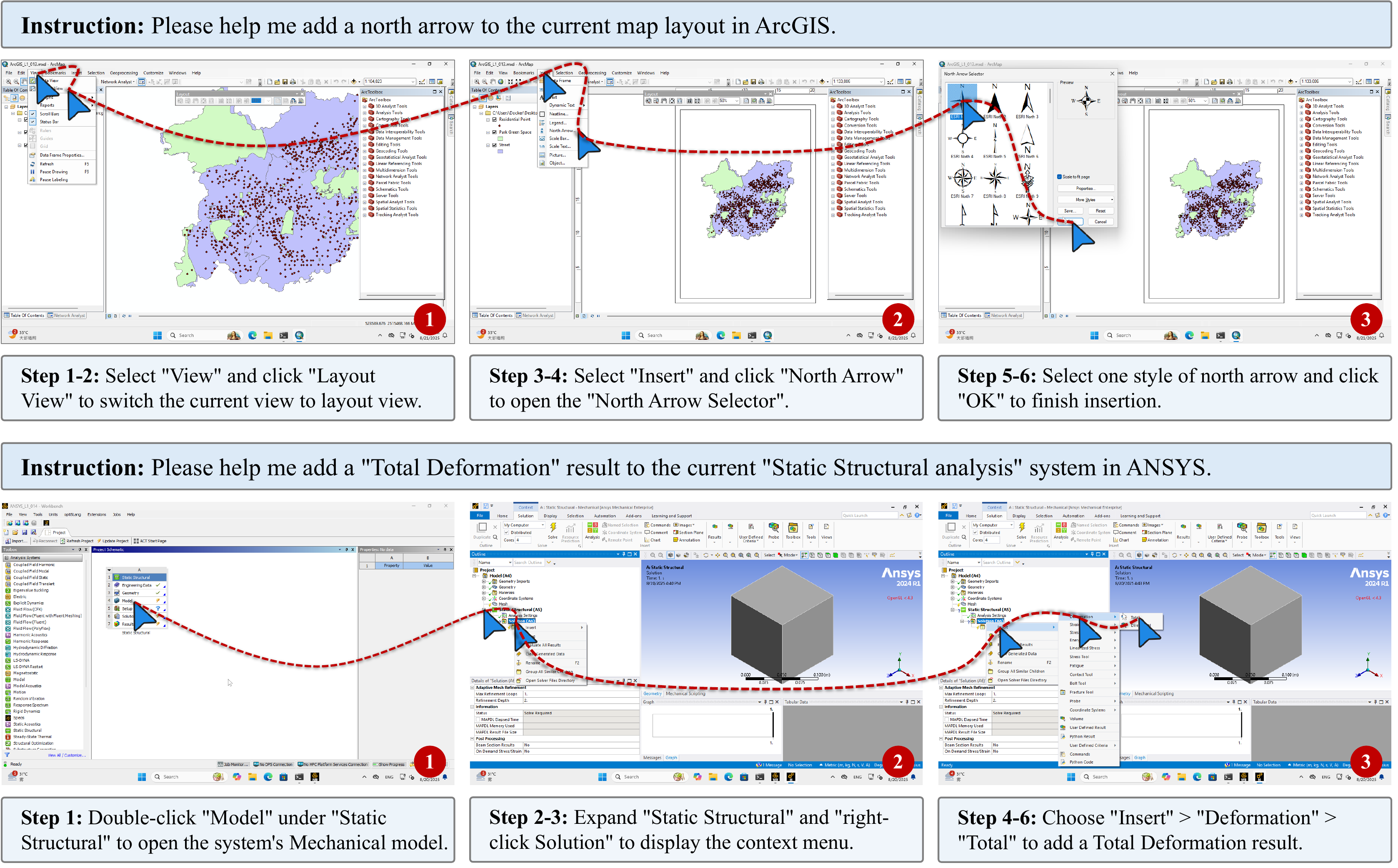}
  \caption{Qualitative Examples of Agent Trajectories.
  The top row shows the process of adding a north arrow in ArcGIS, involving view switching and dialog interaction. 
  The bottom row illustrates adding a simulation result in ANSYS through tree navigation and context menus. 
  Red dashed lines and blue cursors indicate mouse movements and clicks, respectively.}
  \label{fig:agent_trajectories}
\end{figure*}
\section{ProSoftArena Platform}
\label{sec:platform}

\subsection{Executable Professional Software Environment}
\label{subsec:environment}
ProSoftArena is built upon a realistic, interactive computer environment that leverages virtual machine (VM) technology to host a fully functional Windows 11 system, managed via Docker containers. 
We choose Windows as the primary operating system because the majority of complex, industry-standard professional software relies on deep system integration and is either exclusively available or most stable on Windows platform. 
The VM-based design provides an isolated, secure, and clean environment that prevents agents from causing irreversible damage to the host system. 
The snapshot functionality allows efficient, deterministic resets to a pristine state before each task, guaranteeing consistent initial conditions. 
The environment is pre-configured with 13 core professional applications, including Adobe Illustrator, PhotoShop, ImageJ, ChemDraw, R (RGui), Excel, VSCode, NVIVO, ArcGIS, ANSYS, MultiSim, AutoCAD and Solidworks.
These were selected as the most common tools spanning 6 major disciplines and 20 specialized subfields, with 
all software fixed to specific versions to ensure a consistent and reproducible evaluation baseline. 
Beyond evaluation, our environment can also be used as a training platform for agents to learn professional software usage. 
Further implementation details and software specifications are provided in Appendix~\ref{sec:supp_environment}.

\subsection{Automated Evaluation framework}

\noindent
\textbf{Initial Task Environment Setup.}
In real-world scenarios, task requests typically occur in the middle of a workflow, such as editing an open design draft or extending an ongoing data analysis. To faithfully simulate these contextual working states, we design a custom initialization script for each task that performs the following procedures: (i) restore the virtual machine to a pristine state from a clean snapshot; (ii) download required task files from cloud storage to the VM; (iii) launch the specific professional application and load the target files; (iv) execute preprocessing steps, such as adjusting window layout or navigating to specific tool panel, to establish the intended initial context. 
As shown in Figure~\ref{fig:method}, our platform automatically executes the initialization script before each task, ensuring consistent environment setup and reproducible evaluation. 

\noindent
\textbf{Agent-Environment Interaction.}
After environment initialization, the agent engages in an interactive loop with the professional software environment, as illustrated in Figure~\ref{fig:method}. This process can be formalized as a partially observable Markov decision process (POMDP) defined by the tuple \(⟨g, S, A, O, T ⟩\), where \(g\) denotes the user instruction, \(T : S \times A \rightarrow S\) denotes the state transition function, \(S\), \(A\) and \(O\) correspond to the state, action and observation spaces, respectively.
The process repeats until the agent outputs a termination action (DONE or FAIL) or when the maximum number of steps is reached.
In our settings, MLLMs act as the policy model to drive the agent's decision-making process, enabling task planning and action prediction in professional software environments.
Figure~\ref{fig:agent_trajectories} visualizes representative execution trajectories of
an agent interacting with ProSoftArena environment and completing tasks.

\noindent
\textbf{Execution-based evaluation.}
To ensure accurate and reliable assessment of agent performance in professional software environments, we adopt execution-based evaluation that verifies task success by examining final system states and output artifacts, aligning with real-world expectations for professional work quality.
First, we design dedicated evaluation functions for each task according to its requirements and success criteria. 
For example, in PhotoShop image editing tasks, we compare the final output with the expected result using Mean Squared Error (MSE) with predetermined thresholds; for environment configuration tasks in VSCode, we read the user-level configuration files to ensure target settings such as theme and auto-save are correctly applied.
Next, we define task-specific evaluation scripts that execute the following steps: (i) perform post-processing such as activating the target window or saving the current working files; (ii) retrieve relevant output files or application state information from the VM; (iii) execute the evaluation function to determine whether the task objectives have been functionally achieved, returning 1.0 for success or 0.0 for failure.
The entire evaluation is automatically executed, ensuring consistency and reproducibility.
Detailed examples are provided in Appendix~\ref{sec:supp_evaluation_framework}.

\subsection{Human-in-the-loop Evaluation}
\label{subsec:hitl}
We recognize human-agent collaboration as a critical direction for developing practical AI assistants, especially in complex, high-stakes scenarios involving professional software. To gain insights into the agent’s behavior and interaction efficiency in real collaborative settings, we implement a structured human-in-the-loop evaluation framework with two complementary interaction modes.

\noindent
\textbf{Mode 1:} Human-Initiated Takeover. In this mode, a human expert monitors the agent's execution in real-time. When critical errors, significant trajectory deviations, or potentially irreversible actions are detected, the human expert can immediately take control and perform necessary corrective actions in the environment. After this, control is handed back to the agent, which continues its task based on the updated environment.

\noindent
\textbf{Mode 2:} Agent-Initiated Assistance. In this mode, we provide the agent with an additional tool \(ASK\_ACTION\). When the agent is uncertain about the next steps or requires additional information, it can invoke this action, specifying the reason for the request and the exact assistance needed. The human expert then performs the requested action in the environment
and agent proceeds from the updated state.

\section{ProSoftArena Benchmark}
\label{sec:benchmark}


\subsection{Domain and Software Coverage}
\label{subsec:domain}
To comprehensively evaluate multimodal agents in authentic professional software environments, ProSoftArena spans 6 major disciplines, 20 subfields, and 13 representative professional applications. These tools were selected based on their prevalence and essential role in real-world workflows.
Our benchmark surpasses existing ones in both the breadth of domains and the diversity of professional software, providing a comprehensive testbed for assessing agent capabilities across the full spectrum of high-value professional work.
More information about software version and domain examples are provided in Appendix~\ref{sec:supp_environment} and \ref{sec:supp_domain}.

\subsection{Task Annotation}
\label{subsec:task_anno}
To construct realistic and diverse professional software tasks, we engage 12 domain experts proficient in the target applications. 
The full pipeline includes hierarchical task collection, metadata annotation, quality control, and script implementation and validation.

\noindent
\textbf{Hierarchical Task Collection.}
Experts derive L1 tasks from official software manuals, formulate L2 tasks from routine professional workflows and public tutorials (e.g., official software tutorials and YouTube instructional videos), and collaboratively design multi-application L3 pipelines. L4 tasks are created from curated human creative works. When constructing higher-level tasks, annotators identify the underlying L1 operations and ensure any missing ones are added to maintain
coverage across capability levels.

\noindent
\textbf{Metadata Annotation.}
All tasks are annotated within a standardized software environment with fixed versions. Each task includes detailed natural-language instructions, input files, difficulty level, expected outputs, evaluation criteria, a demonstration trajectory, and human execution statistics. 
A full list of metadata fields is provided in Appendix~\ref{sec:supp_metadata}.

\noindent
\textbf{Quality Control.}
Every task is independently reviewed by at least two domain experts to verify feasibility, clarity, and correctness of the expected outputs and evaluation rules. Tasks that are ambiguous, ill-defined, or not reliably evaluable are removed from the benchmark. Additional details of the quality control protocol are provided in Appendix~\ref{sec:supp_quality}.

\noindent
\textbf{Script Implementation and Validation.}
For each task, we implement initialization and evaluation scripts to enable fully automated configuration and scoring. Scripts are validated by reconstructing the intended starting state, having an expert complete the task, and confirming that the evaluation function assigns the correct score. 

\begin{table}[t]
\centering
\caption{Statistics of ProSoftArena}
\resizebox{0.45\textwidth}{!}{
\begin{tabular}{lc}
\toprule
\textbf{Statistic} & \textbf{Number} \\

\midrule
\textbf{Operation Level (L1)} & 252 (57.8\%) \\
- Simple / Middle / Hard & 173 / 67 / 12 \\
- Avg. / Min. / Max. Human Exec. Steps & 5.1 / 2 / 18 \\
- Avg. / Min. / Max. Human Exec. Time (s) & 14.8 / 3 / 65 \\
\midrule
\textbf{Software Level (L2)} & 164 (37.6\%)\\
- Simple / Middle / Hard & 48 / 86 / 30 \\
- Avg. / Min. / Max. Human Exec. Steps & 20.4 / 3 / 82 \\
- Avg. / Min. / Max. Human Exec. Time (s) & 83.1 / 5 / 600\\
\midrule
\textbf{Pipeline Level (L3)} & 10 (2.3\%)\\
- Simple / Middle / Hard & 0 / 0 / 10 \\
- Avg. / Min. / Max. Human Exec. Steps & 86.9 / 46 / 109\\
- Avg. / Min. / Max. Human Exec. Time (s) & 506.8 / 350 / 647 \\
\midrule
\textbf{Creative Level (L4)} & 10 (2.3\%)\\
- Simple / Middle / Hard & 0 / 0 / 10 \\
\midrule
\textbf{Total} & 436 \\ 
- Disciplines / Subfields / Applications & 6 / 20 / 13 \\
- Avg. / Min. / Max. Human Exec. Steps & 12.9 / 2 / 109 \\
- Avg. / Min. / Max. Human Exec. Time (s) & 52.6 / 3 / 647 \\
\bottomrule
\end{tabular}
}
\label{tab:statistics}
\vspace{-7.2mm}
\end{table}
\begin{table*}
  \centering
  \caption{Main Results on L1, L2 and L3 tasks.}
  \resizebox{\textwidth}{!}{
    \setlength{\tabcolsep}{6.9pt}
      \begin{tabular}{ll *{7}{c@{\hspace{6.8pt}}c} c}
        \toprule

        \multirow{3}{*}[-1ex]{\makecell[c]{\textbf{Inputs}}} &
        \multirow{3}{*}[-1ex]{\makecell[c]{\textbf{Model}}} &
        \multicolumn{15}{c}{\textbf{Success Rate (\%) \(\uparrow\)}} \\ 
        \cmidrule(){3-17} 
        && \multicolumn{2}{c}{Art} &
        \multicolumn{2}{c}{Business} &
        \multicolumn{2}{c}{Nature Sci} &
        \multicolumn{2}{c}{Medicine} &
        \multicolumn{2}{c}{Social Sci} &
        \multicolumn{2}{c}{T \& E} &
        \multicolumn{2}{c}{\textbf{Overall}} &
        \multirow{2}{*}{\textbf{L3}} \\
        \cmidrule(){3-16} 
        & & L1 & L2
        & L1 & L2
        & L1 & L2
        & L1 & L2
        & L1 & L2
        & L1 & L2
        & L1 & L2 \\
        \midrule
        \multirow{5}{*}{Screenshot} 
        & \cellcolor{blue}GPT-4o
        & 2.9 & 0.0 & 15.4 & 3.7 & 1.8 & 0.0 & 2.6 & 0.0 & 0.0 & 0.0 & 14.3 & 4.8 & 7.1 & 1.9 & 0.0 \\
        & \cellcolor{blue}o3
        & 28.9 & 0.0 & 37.5 & 22.2 & 22.2 & 0.0 & 10.3 & 0.0 & 0.0 & 0.0 & 24.1 & 9.5 & 20.3 & 6.2 & 0.0 \\
        & \cellcolor{blue}GPT-5
        & 43.6 & 3.6 & 62.5 & 29.6 & 21.4 & 0.0 & 10.3 & 7.1 & 0.0 & 0.0 & 23.1 & 7.1 & 28.9 & 8.6 & 0.0 \\
        & \cellcolor{green}GLM-4.5V
        & 0.0 & 0.0 & 5.1 & 0.0 & 0.0 & 0.0 & 2.6 & 0.0 & 0.0 & 0.0 & 12.5 & 2.4 & 4.1 & 0.6 & 0.0 \\
        & \cellcolor{green}Qwen2.5-VL
        & 8.8 & 0.0 & 20.5 & 7.4 & 5.4 & 0.0 & 5.1 & 0.0 & 0.0 & 0.0 & 19.6 & 2.4 & 11.2 & 1.9 & 0.0 \\
        \midrule
        \multirow{5}{*}{\makecell{Screenshot\\+A11y tree}}
        & \cellcolor{blue}GPT-4o
        & 3.0 & 0.0 & 38.5 & 3.7 & 22.2 & 0.0 & 17.9 & 0.0 & 0.0 & 0.0 & 14.5 & 4.8 & 18.2 & 1.9 & 0.0 \\
        & \cellcolor{blue}o3
        & 34.3 & 3.6 & 60.0 & 25.0 & 27.8 & 0.0 & 23.1 & 7.1 & 6.3 & 0.0 & 39.3 & 7.5 & 34.6 & 7.7 & 0.0 \\
        & \cellcolor{blue}GPT-5
        & 43.6 & 7.1 & 77.5 & 43.5 & 48.1 & 4.0 & 28.2 & 10.7 & 6.3 & 0.0 & 42.9 & 12.2 & 45.1 & 13.5 & 0.0 \\
        & \cellcolor{green}GLM-4.5V
        & 0.0 & 0.0 & 7.7 & 0.0 & 1.8 & 0.0 & 5.1 & 0.0 & 0.0 & 0.0 & 14.3 & 2.4 & 5.9 & 0.6 & 0.0 \\
        & \cellcolor{green}Qwen2.5-VL
        & 10.8 & 0.0 & 33.3 & 18.5 & 16.4 & 0.0 & 5.1 & 0.0 & 0.0 & 0.0 & 23.2 & 4.8 & 16.9 & 4.3 & 0.0 \\
        \midrule
        \multirow{6}{*}{Set-of-Mark}
        & \cellcolor{blue}GPT-4o
        & 0.0 & 0.0 & 23.1 & 14.8 & 7.3 & 0.0 & 2.6 & 0.0 & 0.0 & 0.0 & 16.1 & 2.4 & 9.9 & 3.1 & 0.0 \\
        & \cellcolor{blue}o3
        & 23.7 & 3.6 & 42.5 & 51.9 & 23.6 & 7.7 & 5.1 & 7.1 & 12.5 & 0.0 & 28.6 & 21.4 & 24.2 & 17.9 & 0.0 \\
        & \cellcolor{blue}GPT-5
        & 28.8 & 3.6 & 70.0 & 48.1 & 29.1 & 7.7 & 17.9 & 7.1 & 18.8 & 0.0 & 42.9 & 30.9 & 36.4 & 19.1 & 0.0 \\
        & \cellcolor{green}GLM-4.5V
        & 0.0 & 0.0 & 10.3 & 11.1 & 3.6 & 0.0 & 2.6 & 0.0 & 0.0 & 0.0 & 14.3 & 2.4 & 6.2 & 2.5 & 0.0 \\
        & \cellcolor{green}Qwen2.5-VL
        & 9.9 & 0.0 & 23.1 & 22.2 & 7.3 & 0.0 & 5.1 & 0.0 & 6.3 & 0.0 & 17.9 & 4.8 & 10.7 & 4.9 & 0.0 \\
        \midrule
        \multicolumn{2}{c}{\cellcolor{yellow}Claude 4 Computer Use}
        & 34.6 & 10.7 & 74.4 & 53.2 & 47.1 & 19.2 & 33.3 & 10.7 & 43.8 & 9.1 & 41.7 & 17.5 & 45.9 & 24.4 & 0.0 \\
        \bottomrule
      \end{tabular}
    }
  \label{tab:main}
\end{table*}
\subsection{Statistics}
\label{subsec:statistics}
The statistics of ProSoftArena are presented in Table~\ref{tab:statistics}. Our benchmark comprises 436 tasks spanning 6 disciplines, 20 subfields, and 13 professional software applications, providing broad coverage at considerable scale. Tasks follow the defined capability hierarchy: L1 and L2 constitute the majority, while a small set of challenging L3 and L4 tasks extends the evaluation frontier. As the capability level increases, task difficulty shifts from simple to hard, accompanied by a significant growth in human execution steps and time—from an average of 5.1 steps and 14.8 seconds at L1 to 86.9 steps and 506.8 seconds at L3. ProSoftArena surpass prior benchmarks in task complexity and duration especially at L2 and L3, highlighting the distinctive challenges posed by professional software environments. Further detailed breakdowns are provided in Appendix~\ref{sec:supp_statistics}.

\section{Experiments}
\label{sec:exp}
We evaluate three categories of agent backbones: proprietary MLLMs (GPT-4o, GPT-5, o3), open-source MLLMs (Qwen2.5-VL-72B-Instruct, GLM-4.5V), and the specialized computer-use agent Claude 4 Sonnet. For MLLM-based agents, we provide three observation types: (i) Screenshot: raw screen capture; (ii) *Screenshot+A11y*: screenshot combined with simplified accessibility tree; (iii) Set-of-Marks (SoM): annotated screenshot with labeled interactable elements. Claude 4 is uniquely granted direct environmental control, autonomously deciding when and how to gather observations. For L1–L3 tasks, we use success rate (SR) as the primary metric. L4 tasks undergo subjective evaluation by comparing agent outputs with human artifacts, with detailed results provided in Appendix~\ref{sec:supp_l4}.

\subsection{Main Results}
Overall performance across models and capability levels is summarized in Table~\ref{tab:main}. 

\noindent
\textbf{Performance across capability levels.} Despite being the easiest tier, L1 still has substantial headroom, with the best model remaining below 50\% success. L2 is the primary bottleneck. The sharp drop from L1 to L2 reveals a clear gap between basic GUI manipulation and software-level usage, which demands domain knowledge and multi-step decision-making. Although agents can execute atomic operations, they struggle to compose these into coherent, parameterized sequences that complete software-specific tasks. Cross-application workflows (L3) remain largely out of reach in professional software settings, where long-horizon planning, consistent state tracking, and semantic alignment across professional applications are the critical challenges.

\noindent
\textbf{Performance gap between base models. }
Closed-source MLLMs consistently lead across all settings, with the latest-generation GPT-5 demonstrates particularly strong performance. This suggests that stronger backbones confers superior capabilities in long-horizon planning, stable GUI manipulation and deeper intrinsic understanding of professional software functionality. We further observe that additional textual information (a11y tree) yields larger gains for stronger backbones, underscoring their enhanced ability to  fuse and reason over multimodal inputs. Within the open-source cohort, Qwen2.5-VL performs best, reaching and locally surpassing GPT-4o; we attribute this to its training-enhanced GUI ability, critical for complex, dense interfaces in professional software. Finally, the specialized computer-use agent, Claude, achieves the overall best performance, indicating that agents explicitly endowed with strong UI-operation routines and tool-use priors are currently the most compatible with professional software environments.

\noindent
\textbf{Impacts of different Observations.}
Richer observations are necessary for professional software environments: complex, dense GUIs make Screenshot-only inputs brittle.  Augmenting with either accessibility tree (a11y) or Set-of-Marks (SoM) improves grounding by constraining the search space and exposing actionable structure. 
Importantly, there is no global winner between a11y and SoM in our settings; each exhibits distinct sweet spots and failure modes tied to software characteristics. Screenshot+A11y is preferable when: (i) controls are tiny and tightly packed. SoM labels can occlude adjacent elements and impair interaction accuracy, while Screenshot+a11y avoids marking on-screen marking and still enables stable clicking via candidate coordinates from the a11y tree; (ii) some controls lack exposed coordinates. SoM cannot annotate these elements, and agents under SoM settings tend to predict labels rather than issue absolute coordinates, leading to “visible-but-unclickable” failures. In contrast, Screenshot+a11y retains both pixel-space clicks and candidate coordinate pathways, ensuring more robust interaction. SoM excels when: (i) toolbars/panels are icon-heavy and text-sparse. The a11y tree may indicate that "an interactive element exists here" but fails to convey the semantic meaning of the icon. SoM, by exposing the icon within a bounding box, allows the model to merge recognition and localization into a single decision; (ii) spacing between controls is sufficient. 

\begin{figure}[!t]
\centering

\begin{minipage}{0.48\linewidth}
  \centering
  \includegraphics[width=\linewidth]{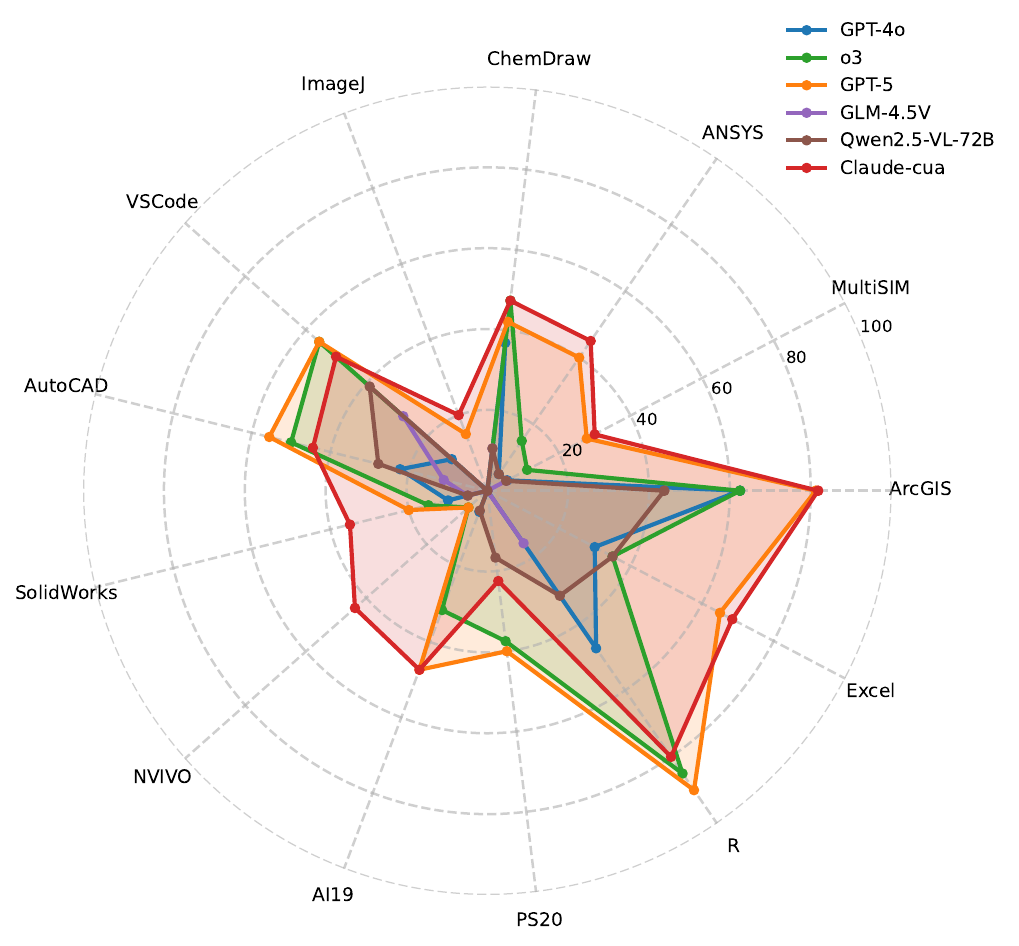}
\end{minipage}
\hfill
\begin{minipage}{0.48\linewidth}
  \centering
  \includegraphics[width=\linewidth]{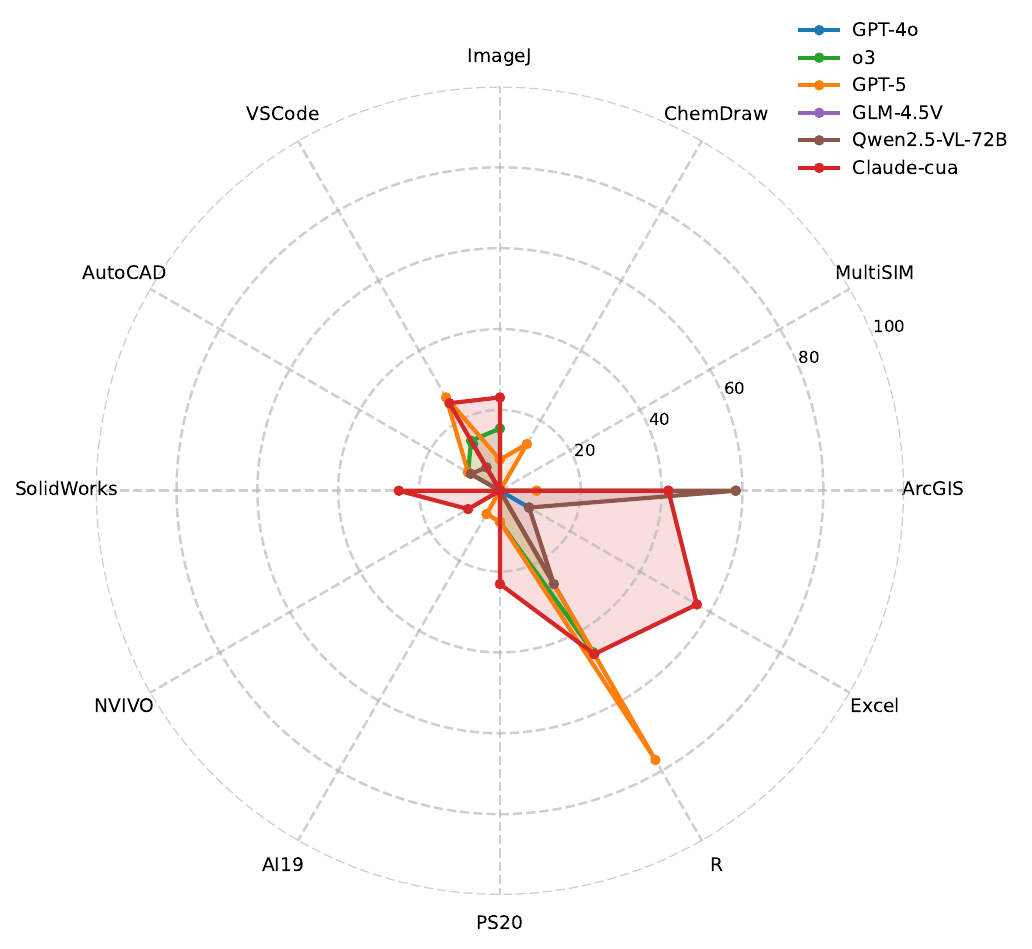}
\end{minipage}

\caption{Model performance across applications. Left: L1 tasks. Right: L2 tasks. Both under Screenshot+A11y observations.}
\vspace{-1.5em}
\label{fig:radar}
\end{figure}

\noindent
\textbf{Performance across domains.} 
We observe substantial variation across domains, closely tied to each field's task characteristics and core software. Business shows higher success rates primarily because its core application, R, has a relatively simple GUI and supports script-based task completion, reducing GUI interaction and allowing models focus on instruction understanding and code generation. By contrast, Social Science underperforms, as tasks require prior knowledge of qualitative analysis, and the core application NVivo involves extensive cross-file operations such as format conversions and data imports/exports, posing challenges to current agents. 
We provide a fine-grained analysis of each model’s performance on individual software in Figure~\ref{fig:radar}. Poorly performing software generally falls into three categories: (i) those requiring strong visual reasoning capabilities, such as Art \& Design applications, which present dense and complex GUIs while also demanding comprehension and reasoning about visual contents; (ii) those relying heavily on domain-specific knowledge, such as Health \& Medicine applications; and (iii) those demanding frequent, precise GUI interactions, such as MultiSim. Addressing these challenges require future work to focus on enhancing complex visual grounding, long-horizon action planning, and domain knowledge integration.

\subsection{Error Analysis}
Based on our fine-grained analysis of 150 failed trajectories across different models and domains, we identify several predominant error patterns that hinder agents' performance in professional software environments. 
\textbf{Task Planning Errors} are one of the major failure modes, where agents misinterprete task instructions, predict incorrect action sequences, omits crucial operations, or selects unsuitable workflows.
\textbf{Domain and Tool Knowledge Gaps} constitute another major limitation, where agents misunderstand domain-specific terminology in instructions, fail to locate critical features within application interfaces, or confuse the functionality of different UI elements.
\textbf{Visual Grounding Inaccuracies} present persistent execution-level challenges, where agents mispredict coordinates or labels for target UI elements despite generating plausible action plans.
Additional systematic errors include predicting invalid actions outside the predefined action space
and repeating operations even after successful execution, indicating failures in action history tracking or UI state recognition.
See Appendix~\ref{sec:supp_failure} for detailed failure cases.

\subsection{Ablation Study}
Building on our error analysis, which identified critical agent limitations in visual grounding, domain knowledge, and action history tracking, we conduct targeted ablations to investigate these challenges and derive design insights.

\begin{table}[t]
\centering
\caption{Ablation on Visual Grounding Inputs.}
\resizebox{0.5\textwidth}{!}{
\setlength{\tabcolsep}{6pt}
\begin{tabular}{l 
*{3}{c@{\hspace{5.5pt}}c} c}
\toprule
\multirow{2}{*}[-0.5ex]{\makecell[c]{\textbf{Observation}}} & \multicolumn{2}{c}{\textbf{SR (\%)}} & \multicolumn{2}{c}{\textbf{Time (s)}} & \multicolumn{2}{c}{\textbf{Cost (Tokens)}} \\
\cmidrule{2-3}
\cmidrule{4-5}
\cmidrule{6-7}
 & L1 & L2 & L1 & L2 & L1 & L2 \\
\midrule
Som from A11y & 0.0 & 0.0 & 1146.3 & 2899.1 & 308.2k & 369.4k \\
Som from A11y+Omni & 10.5 & 0.0 & 1761.2 & 5764.2 & 414.1k & 750.4k \\
\midrule
Som+Screenshot & 21.1 & 6.7 & 1553.8 & 4276.2 & 416.2k & 766.3k \\
Som+Screen+A11y & 5.3 & 0.0 & 1458.9 & 3614.9
 & 616.1k & 1031.8 \\
\bottomrule
\end{tabular}
}
\label{tab:abla_visual}
\end{table}

\noindent
\textbf{Ablation on Visual Grounding Inputs.} We investigate where richer inputs can improve visual grounding. As shown in Table~\ref{tab:abla_visual}, we evaluate four observation configurations on Illustrator with Qwen2.5-VL: (i) baseline SoM constructed from accessibility tree; (ii) enhanced SoM combining accessibility tree with detections from OmniParser-v2, a powerfull, specialized model for visual UI parsing; (iii) SoM combined with raw screenshot; and (iv) a full integration of SoM, screenshot, and the complete A11y tree. Adding additional information yields performance gains but also increases cost, especially on L2 tasks that require more execution steps. The effect is not monotonic: adding both the screenshot and the a11y tree on top of SoM leads to a clear performance drop, suggesting that overloaded inputs exceed the model’s effective processing capacity and introduce conflicting cues. "SoM+Screenshot" proves most effective, achieving a 21.1\% performance improvement for L1 tasks with acceptable cost increases, whereas for more complex L2 tasks, the substantial cost outweighs the limited gains, making it economically inefficient.

\begin{figure}[!t]
	\centering
    \includegraphics[width=0.9\linewidth]{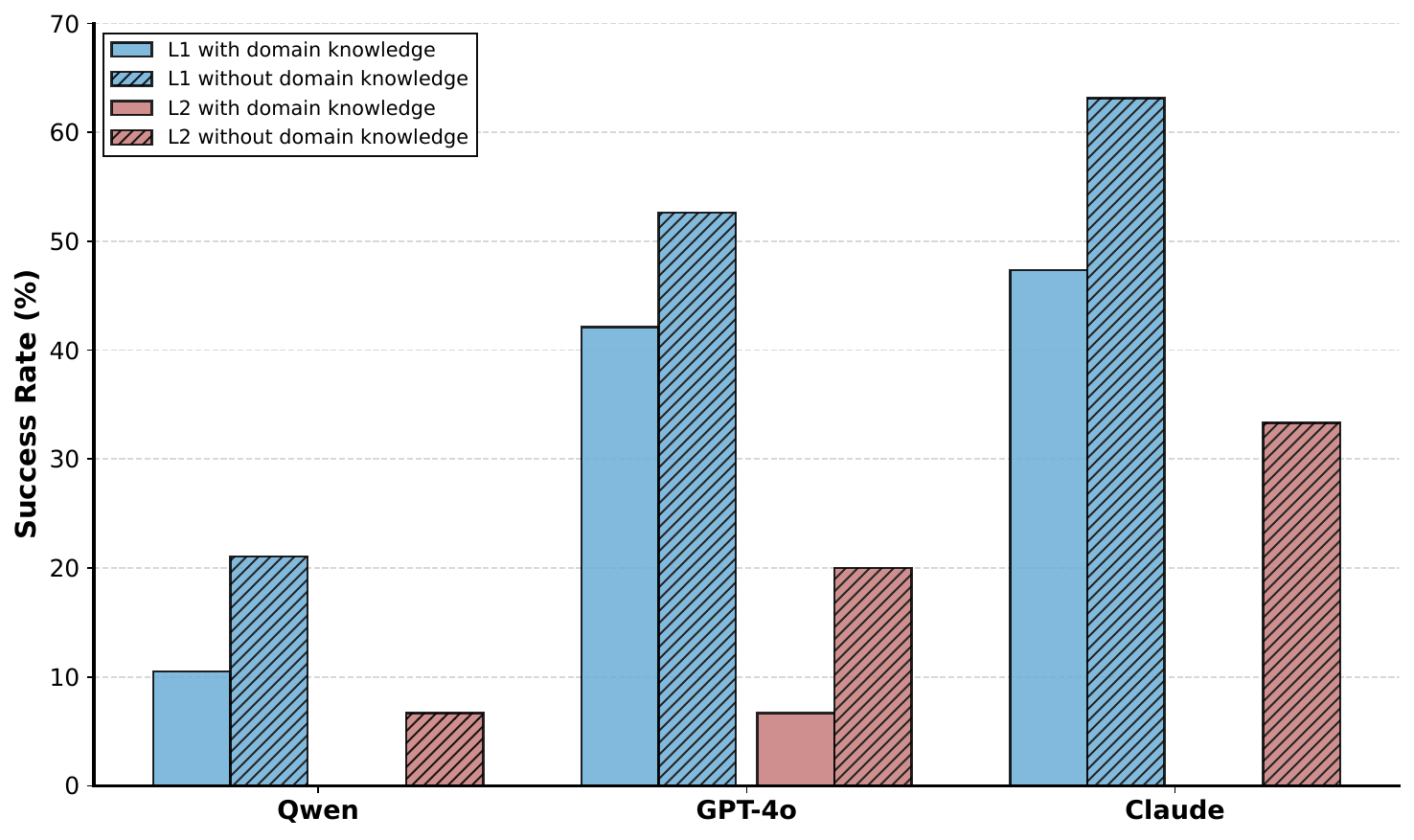}
    \vspace{-0.5em}
	\caption{Success rates of Qwen, GPT-4o, and Claude on L1 and L2 tasks with and without domain knowledge.}
	\label{fig:zhuzhuang}
    \vspace{-3mm}
\end{figure}

\noindent
\textbf{Ablation on Domain Knowledge.} We examine whether
domain priors improve execution on ChemDraw by prepending compact “knowledge cards” that specify molecular formulas, and salient structural constraints. Results are shown in Figure~\ref{fig:zhuzhuang}. We observe consistent gains across models, with larger improvements on L2, where tasks depend more heavily on domain knowledge. Among the models, Claude benefits the most, while Qwen’s improvement is comparatively limited, indicating its primary bottleneck lies in multimodal interaction rather than knowledge comprehension. Overall, injecting domain knowledge is a practical way to boost performance in professional tools.

\begin{figure}[!t]
	\centering
    \includegraphics[width=0.9\linewidth]{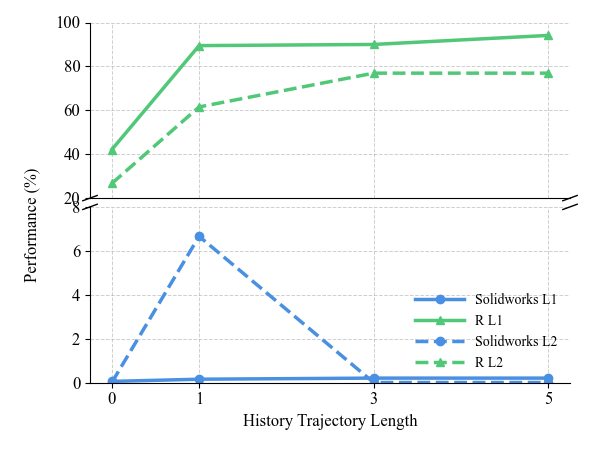}
    \vspace{-1em}
	\caption{Performance (\%) across different history trajectory lengths for SolidWorks and R tasks at Level 1 and Level 2.}
	\label{fig:tiaoxing}
    \vspace{-3mm}
\end{figure}

\noindent
\textbf{Ablation on Action History Length.} We examine how action history length affects task performance by varying the number of historical steps (0, 1, 3, 5) provided to GPT-5. As shown in Figure~\ref{fig:tiaoxing}, longer history contexts consistently improve success rates, particularly in R, as recalling recent edits helps the agent chain script operations and recover from minor detours. Performance saturates beyond 3 steps, suggesting diminishing returns. SolidWorks shows only modest improvement, as its rapidly changing, geometry-centric GUI makes earlier references to coordinates and tool modes stale. In practice, short, recent histories are beneficial for text/script-centric workflows, whereas complex environments like CAD demand more sophisticated state tracking beyond simple history concatenation.

\subsection{Human-in-the-loop Evaluation}
\begin{table}[t]
\centering
\caption{Human-in-the-loop evalution results.}
\resizebox{0.48\textwidth}{!}{
\begin{tabular}{lllcccc}
\toprule
Model & Level & \textbf{Modes} & \textbf{SR (\%)} & \textbf{Steps} &\textbf{HIC}& \textbf{HIT (s)}\\
\midrule
\multirow{6}{*}{GPT-5} & 
 & w/o H & 55.6 & 8.2 & - & - \\
&L1 & w HIT & 100.0 & 7.5 & 1.0 & 16.0 \\
& & w AIA & 55.6 & 8.9 & 0 & 0 \\
\cmidrule(){2-7} 
 &
 & w/o H & 26.7 & 34.2 & - & - \\
&L2 & w HIT & 73.3 & 19.8 & 2.9 & 47.5 \\
& & w AIA & 33.3 & 32.6 & 0.3 & 2.7 \\
\midrule
\multirow{6}{*}{Qwen2.5-VL} & 
 & w/o H & 38.9 & 17.0 & - & - \\
&L1 & w HIT & 100.0 & 10.7 & 1.4 & 20.8 \\
& & w AIA & 38.9 & 19.0 & 0 & 0 \\
\cmidrule(){2-7} 
 &
 & w/o H & 6.7 & 44.6 & - & - \\
&L2 & w HIT & 66.7 & 12.5 & 3.9 & 62.4 \\
& & w AIA & 13.3 & 42.3 & 0.2 & 2.6 \\
\bottomrule
\end{tabular}
}
\label{tab:HITL}
\vspace{-1.5em}
\end{table}


We conduct evalution on VSCode with GPT-5 and Qwen2.5-VL, comparing two interaction modes Human-Initiated Takeover (HIT) and Agent-Initiated Assistance (AIA) against the fully autonomous setting (w/o H). Results are summarized in Table~\ref{tab:HITL}, where steps, HIC, and HIT denote the average execution steps, human intervention count, and human operation time (s) per task, respectively.

HIT consistently enhances success rates and efficiency across models and levels, with larger gains on harder tasks and for the weaker model. For example, with HIT, Qwen2.5-VL's success rate on L2 tasks increased from 6.7\% to 66.7\%, while the average number of steps decreased from 44.6 to 12.5. This demonstrates that targeted human intervention can effectively correct agent errors, reduce inefficient exploration, and optimize the overall execution path. By contrast, AIA provides limited gains, primarily due to the low frequency of help requests issued by the agents. This suggests that current models lack calibrated self-assessment and struggle to proactively seek assistance under uncertainty. Overall, human-agent collaboration shows substantial practical value in professional software environments, where modest human oversight can greatly improve system reliability. Cultivating accurate self-awareness and proactive collaboration in agents remains critical for efficient human-in-the-loop systems.


\section{Conclusion}
We propose ProSoftArena, the first comprehensive benchmark and platform specifically for evaluating multimodal agents in professional software environments.
Our work establishes the first hierarchical taxonomy of agent capabilities in professional software, providing a systematic framework for capability probing and a roadmap for future research.  We also incorporate a unique human-in-the-loop evaluation paradigm, assessing agent's collaborative efficiency beyond mere autonomous success. 
Through extensive experiments, we identify challenges in current agents and provide insights for efficient design principles. 
Our work lays a foundation for advancing multimodal agents in professional software scenarios.


\bibliography{main}
\bibliographystyle{icml2026}

\newpage
\appendix
\section{ProSoftArena Platform: Additional Details}
\label{sec:supp_platform}

\subsection{Executable Professional Software Environment}
\label{sec:supp_environment}
ProSoftArena is constructed upon a realistic and interactive computing framework, utilizing virtual machine (VM) technology to host a fully functional Windows 11 operating system managed via Docker containers. We selected Windows as the primary platform given its dominance in the professional software ecosystem, where complex, industry-standard tools often require deep system integration and are frequently exclusive to or most stable on this OS. The VM-based architecture ensures an isolated and secure execution space, preventing agents from causing irreversible damage to the host system. Crucially, the environment leverages snapshot functionality to enable efficient, deterministic resets to a pristine state before each task, thereby guaranteeing consistent initial conditions for every evaluation.

The environment is pre-configured with 13 core professional applications, including Adobe Illustrator, Photoshop, ImageJ, ChemDraw, R, Excel, VSCode, NVivo, ArcGIS, ANSYS, MultiSim, AutoCAD, and SolidWorks. These applications were rigorously selected to cover representative workflows across 6 major disciplines and 20 specialized subfields. To ensure a consistent and reproducible evaluation baseline, all software is fixed to specific versions. Detailed specifications, including software versions and release builds, are provided in Table~\ref{tab:software_specs}. 

Beyond evaluation, this environment also serves as a scalable training platform, enabling agents to learn and master complex professional software usage.

\begin{table*}[t]
  \centering
  \caption{List of Professional Software, Versions, Disciplines, and Usage Descriptions in ProSoftArena.}
  \label{tab:software_specs}
  \resizebox{\textwidth}{!}{
    \begin{tabular}{l c c c p{7.5cm}}
      \toprule
      \textbf{Software} & \textbf{Version} & \textbf{Discipline} & \textbf{Subfield} & \textbf{Usage} \\
      \midrule
      
      Adobe Photoshop & 26.7 & Art \& Design & Art & Digital image editing and photo retouching. \\
      \midrule
      
      Adobe Illustrator & 29.5.1 & Art \& Design & Design & Vector graphic design and layout print.\\
      \midrule
      
      Microsoft Excel & 2024 & Business, Science & Accounting, Management, Math & Financial modeling, business intelligence, and numerical analysis. \\
      \midrule
      
      R (RGui) & 4.5.0 & \makecell{Business, Science, \\Health \& Medicine} & Finance, Math, Public Health & Statistical computing, econometric modeling, and health policy modeling. \\
      \midrule
      
      ImageJ & 15.4 & Science, Health \& Medicine & \makecell{Biology, Diagnostics \& \\Laboratory Medicine} & Microscopic/medical image analysis, cellular quantification, and diagnostics. \\
      \midrule
      
      ChemDraw & 20.0 & Science, Health \& Medicine & Chemistry, Pharmacy & Molecular structure design, chemical property analysis, and drug design. \\
      \midrule
      
      ArcGIS & 10.2 & Science & Geography & Spatial analysis, cartography, and geographical studies. \\
      \midrule
      
      ANSYS & 2024 R1 & Science, Tech \& Engineering & Physics, Energy Power, Materials & Finite element analysis (structural mechanics, thermal dynamics) and multiphysics simulation. \\
      \midrule
      
      Multisim & 14.3 & Science, Tech \& Engineering & Physics, Electronics & Electronic circuit simulation and PCB design. \\
      \midrule

      NVivo & 20 v1.7.2 & Humanities & Social Science \& Sociology & Systematic coding, thematic organization, and theoretical analysis of qualitative data (e.g., interviews). \\
      \midrule

      VSCode & 1.99.3 & Technology \& Engineering & Computer Science & Integrated development environment (IDE) for programming and software development. \\
      \midrule
      
      AutoCAD & 2026 & Technology \& Engineering & Architecture \&  Engineering & Technical drafting and construction documentation. \\
      \midrule
      
      SolidWorks & 2025 SP1.2 & Technology \& Engineering & Mechanical Engineering & Parametric modeling and assembly design. \\
      
      \bottomrule
    \end{tabular}
  }
\end{table*}

\subsubsection{Observation Space} 
\label{sec:supp_observation}
The observation space \(O\) denotes the subset of the computer environment state \(s_t\in S\) that can be perceived by the agent at each time step. Our environment provides multimodal observations comprising screen representations and additional computer-state signals. We evaluate three types of screen representations: (i) pixel screenshot image that captures the current screen content and window layout; (ii) XML-format accessibility (a11y) tree with structured attributes (control names, bounding boxes, interactable states); and (iii) Set-of-Marks (SoM) overlay that annotates actionable elements with unique identifiers. Visual examples of these observation types are illustrated in Table~\ref{tab:obs_examples}. Beyond the screen, we supply indirect observations including clipboard contents and session metadata (titles of all open windows and the current foreground window).

\subsubsection{Action Space} 
\label{sec:supp_action}
We implement a unified action space \(A\) that encompasses 13 core mouse and keyboard actions covering the full spectrum of human-computer interactions, including mouse movement, clicks, dragging, keystrokes, and hotkeys. 

To support strategic task management, we add three control actions: WAIT (pause to accommodate interface latency), DONE (declare successful completion), and FAIL (indicate infeasibility/abort). Together, these actions enable agents to execute precise low-level manipulations while managing high-level task flow, providing comprehensive control over professional software environments. 
Complete action definitions are listed in Table~\ref{tab:action_space}.

\subsection{Agent Execution Trajectory}
Given a policy \(\pi\), the agent predicts action \(a_t \in A\) at each time step \(t\) based on the goal \(g\) and current observation \(o_t \in O\). 
The professional application then executes \(a_t\) inside the VM and updates the state \(s_t\) to \(s_{t+1}\). Subsequently, the agent obtains an updated observation \(o_{t+1}\) reflecting the new environment state \(s_{t+1}\).
The process repeats until the agent outputs a termination action (DONE or FAIL) or when the maximum number of steps is reached (\(t > t_{max}\)). 


\subsection{Automated Evaluation Framework}
\label{sec:supp_evaluation_framework}

\begin{figure*}[t]
  \centering
  \includegraphics[width=\textwidth]{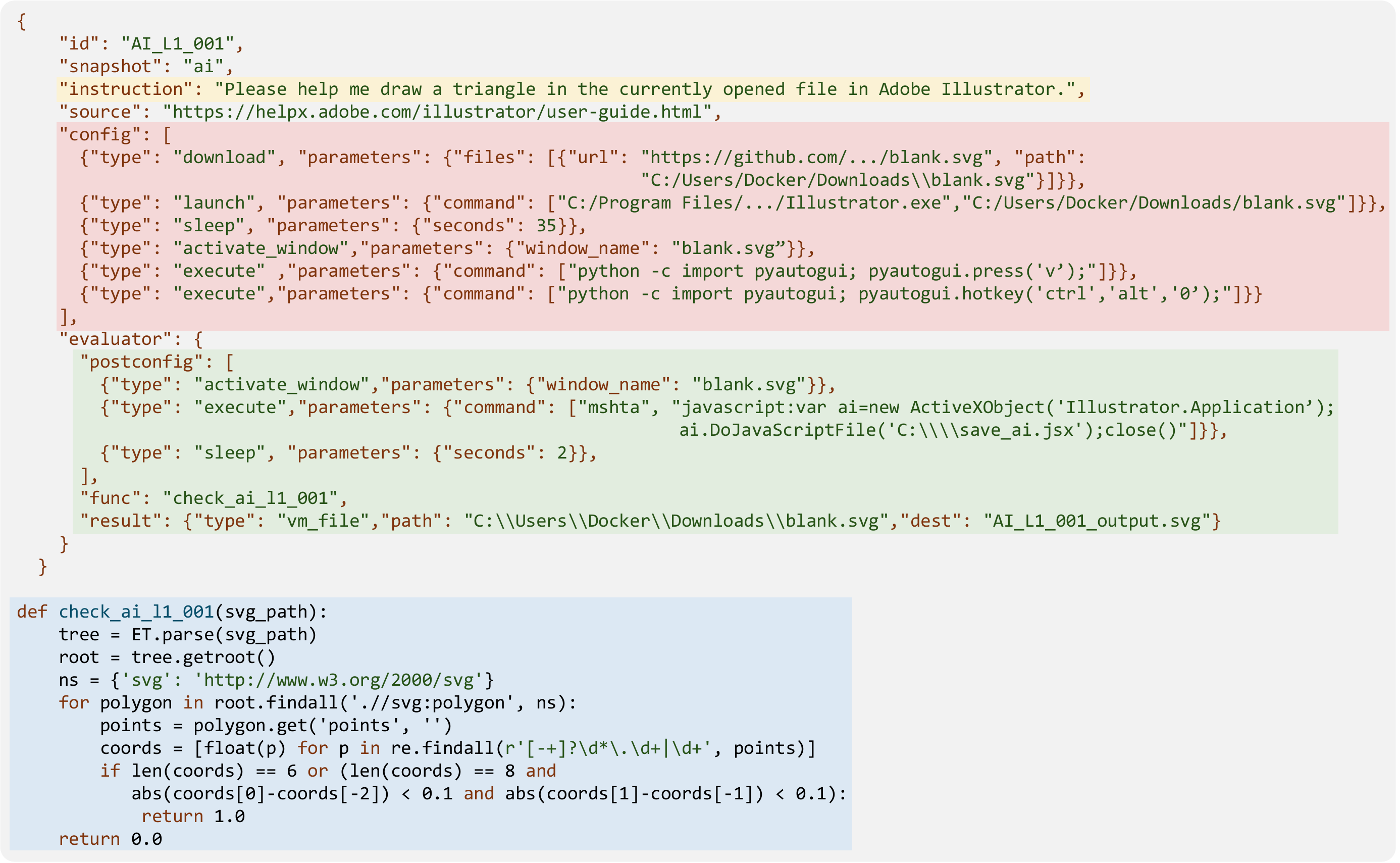}
  \caption{Initialization and Evaluation Script Example for Task "AI\_L1\_001" (Adobe Illustrator).}
  \label{fig:config_script}
\end{figure*}
\textbf{Initial Task Environment Setup.} As shown in the red block of Figure~\ref{fig:config_script}, the initialization script ("config") first downloads the specific task file "blank.svg" from the cloud storage to the VM path "C:/Users/Docker/Downloads/". It then executes the launch command to open Adobe Illustrator and load this file. To standardize the initial context, the script executes a series of preprocessing steps: it activates the target window "blank.svg", uses Python scripts to simulate keyboard shortcuts: 'v' to switch to the Selection Tool and 'ctrl+alt+0' to fit the artboard to the window. This ensures agents start in a consistent, ready-to-work state.

\noindent
\textbf{Execution-based Evaluation.} Upon task completion, the system enters the evaluation phase shown in the green block in Figure~\ref{fig:config_script}. It executes the postconfig sequence to activate the Illustrator window, inject a JavaScript command via mshta to force-save the current canvas content back to the local disk. The system then retrieves the modified file from the VM for verification. Finally, the execution-based evaluation function "check\_ai\_l1\_001" (shown in the blue block in Figure~\ref{fig:config_script}) is called. This function parses the XML structure of the retrieved SVG file and check whether a triangle exists as the task instruction requests.

\section{ProSoftArena Benchmark: Construction Details and Data Samples}
\label{sec:supp_benchmark}
\subsection{Domain and Software Coverage}
\label{sec:supp_domain}
We systematically investigate 6 core disciplines spanning 20 subfields and select 13 representative professional software applications. This selection prioritizes software that is most widely used and fundamental to executing core workflows within their respective domains.

\noindent
\textbf{Art \& Design.} We consider two key subfields in this domain: artistic creation and visual design. For artistic creation, we include Adobe Photoshop, which professionals use for digital image editing and photo retouching. For visual design, we incorporate Adobe Illustrator, dedicated to vector graphic design and print layout. A sample task from this domain is: "Remove all the birds from the opened image in Photoshop, then adjust the brightness of the image to 20 and the vibrance to 100."

\noindent
\textbf{Business.} We investigate three critical subfields: accounting, management, and finance. In accounting and management workflows, Microsoft Excel serves as the primary tool for financial modeling and business intelligence tasks. For advanced financial analysis, we include R, which supports statistical computing and econometric modeling. A representative task requires: "Add Slicers for 'Region' and 'Sales' to the data in range 'A1:B3' in the opened sheet to enable regional sales analysis."

\noindent
\textbf{Science.} This domain covers five core subfields: biology, chemistry, geography, mathematics, and physics. For biology, we include ImageJ, which is widely used for microscopic image analysis and cellular quantification. In chemistry, we incorporate ChemDraw, dedicated to molecular structure design and chemical property analysis. Geographical studies are represented by ArcGIS, which serves as the standard platform for spatial analysis and cartography. Mathematical computing relies on Excel, employed for numerical analysis, and R, specialized in statistical modeling. For physics applications, we include ANSYS for multiphysics simulation and Multisim for electronic circuit design. Representative tasks include: "Measure the angles of four leaf veins from top to bottom in ImageJ and export the results to results.csv," and "Classify the service radius of parks based on park levels using spatial analysis tools in ArcGIS."

\noindent
\textbf{Health \& Medicine.} This domain encompasses the sub-fields of diagnostics \& laboratory medicine, pharmacy, and public health. For the first, we include ImageJ, which supports medical image analysis and diagnostic quantification. In pharmacy, we incorporate ChemDraw, widely used for pharmaceutical molecule analysis and drug design. Public health research is represented by R, utilized for statistical analysis in epidemiological studies and health policy modeling. A sample task in this domain is "Analyze the molecular structure information of Acetylsalicylic acid (Aspirin) using ChemDraw."

\noindent
\textbf{Humanities \& Social Science.} This domain centers on qualitative research methodologies in sociology and related subfields. We incorporate NVivo, the premier software platform in this field, which enables systematic coding, thematic organization, and theoretical analysis of qualitative data such as interview transcripts and survey responses. A representative task requires: "Conduct Matrix Coding query in NVivo to compare perspectives of different respondents across various aspects of 'mental-emotional wellbeing.'"

\noindent
\textbf{Technology \& Engineering.} This domain spans six core subfields with their respective specialized software. For architecture and engineering, we include AutoCAD, employed for technical drafting and construction documentation.  Computer science utilizes VSCode, which provides the integrated environment for programming and software development. For electronics engineering, we incorporate Multisim, dedicated to circuit simulation and PCB design. For energy power and materials science, we include ANSYS, widely used for finite element analysis including structural mechanics and thermal dynamics. Mechanical engineering employs SolidWorks for parametric modeling and assembly design. Representative tasks include: "Draw a standard outdoor badminton court and annotate the dimensions in AutoCAD," and "Model three components of a syringe in SolidWorks: a barrel, a plunger, and a needle."

See Table~\ref{tab:software_specs} for detailed software specifications, versions, and domain mappings.

\subsection{Metadata Annotation}
\label{sec:supp_metadata}
\begin{table}[t]
  \centering
  \caption{Metadata Annotation Example.}
  \label{tab:metadata_example}
  \resizebox{\columnwidth}{!}{
    \setlength{\tabcolsep}{6pt}
    \renewcommand{\arraystretch}{1.1} 
    \begin{tabular}{l p{6.1cm}}
      \toprule
      \textbf{Metadata Field} & \textbf{Content / Value} \\
      \midrule
      \textbf{Task ID} & \texttt{AI\_L1\_001} \\
      \midrule
      \textbf{Instruction} & Please help me draw a triangle in the currently opened file in Adobe Illustrator. \\
      \midrule
      \textbf{Context Resources} & \texttt{blank.svg} (Input file loaded at start) \\
      \midrule
      \textbf{Task Source} & Official User Guide ({ \url{https://helpx.adobe.com/illustrator/user-guide.html}}) \\
      \midrule
      \textbf{Core Operations} & Shape Drawing / Polygon Tool Usage \\
      \midrule
      \textbf{Expected Output} & \texttt{AI\_L1\_001\_output.svg} (Must contain a valid 3-vertex polygon) \\
      \midrule
      \textbf{Difficulty} & Easy \\
      \midrule
      \textbf{Human Reference} & 
      \tabincell{l}{
      - \textit{Demo:} \texttt{AI\_L1\_001.mp4} \\
      - \textit{Steps:} 3 \\
      - \textit{Time:} 10s}
      \\
      \bottomrule
    \end{tabular}
  }
\end{table}
All annotations are conducted within a standardized software environment with fixed application versions to ensure consistency in task execution. 
For each task, domain experts curate a comprehensive metadata profile comprising:
\begin{itemize}
    \item \textbf{Task ID:} A unique identifier for each task.
    \item \textbf{Instruction:} A precise natural language directive specifying the user goal. Instructions are rigorously formulated to ensure the target output is unambiguous and objective.
    \item \textbf{Context Resources:} The necessary input assets to complete the task (e.g., files, datasets or images).
    \item \textbf{Task Source:} The task’s provenance (e.g., manual section, tutorial URL, or expert-authored).
    \item \textbf{Core Operations:} For higher-level tasks (L2/L3), annotators explicitly map the workflow back to the atomic operations (L1) required (e.g., "Shape Drawing"), enabling fine-grained capability analysis.
    \item \textbf{Expected Output:} Defines the tangible artifacts (e.g., saved files, exported charts) or system states used for objective evaluation.
    \item \textbf{Difficulty \& Human Baseline:} Experts provide a subjective difficulty rating (Easy, Medium, Hard) and record their own execution metrics, including a standard demonstration video, total operation steps, and completion time, serving as a "Human Expert" baseline.
\end{itemize}

Table~\ref{tab:metadata_example} presents a concrete example of these metadata fields for an Operation Level task in Adobe Illustrator.

\begin{table*}[t]
  \centering
  \caption{Definition of the Unified Action Space, encompassing 13 core primitive actions for low-level mouse and keyboard interaction and 3 high-level control actions for strategic task management.
  }
  \label{tab:action_space}
  \resizebox{0.7\textwidth}{!}{
    \begin{tabular}{c l l}
      \toprule
      \textbf{Action} & \textbf{Function} & \textbf{Description} \\
      \midrule
      
      \multirow{2}{*}{Mouse Movement} & \texttt{move\_id(id)} & Move cursor to a specific UI element via SoM ID. \\
       & \texttt{move\_abs(x, y)} & Move cursor to absolute coordinates $(x, y)$. \\
      \midrule
      
      \multirow{3}{*}{Clicking} & \texttt{single\_click()} & Perform a single left click. \\
       & \texttt{double\_click()} & Perform a double left click. \\
       & \texttt{right\_click()} & Perform a right click (context menu). \\
      \midrule
      
      Scrolling & \texttt{scroll(direction)} & Scroll the screen in the specified direction. \\
      \midrule
      
      \multirow{2}{*}{Keyboard} & \texttt{write(text)} & Input a text string into the active field. \\
       & \texttt{press(key)} & Press a specific key or hotkey combination. \\
      \midrule
      
      \multirow{3}{*}{Clipboard} & \texttt{copy\_text(text)} & Copy specified text to the system clipboard. \\
       & \texttt{copy\_image(image)} & Copy an image to the system clipboard. \\
       & \texttt{paste()} & Paste content from the clipboard. \\
      \midrule
      
      \multirow{2}{*}{System Control} & \texttt{open\_program(prog)} & Launch a specific professional application. \\
       & \texttt{switch\_to\_app(win)} & Switch focus to a target application window. \\
      \midrule
      
      \multirow{3}{*}{Task Control} & \texttt{WAIT()} & Pause execution to accommodate interface latency. \\
       & \texttt{DONE()} & Declare successful task completion. \\
       & \texttt{FAIL()} & Abort task due to infeasibility or error. \\
      
      \bottomrule
    \end{tabular}
  }
\end{table*}
\subsection{Quality Control}
\label{sec:supp_quality}
Following the data annotation phase, we implement a rigorous quality control protocol to ensure the reliability and validity of the benchmark. Each task is subjected to independent cross-validation by at least two domain experts. The review process strictly assesses the following dimensions:

\begin{itemize}
    \item \textbf{Feasibility: }Verifying that the task is technically completable within the standardized software environment.
    \item \textbf{Instructional Clarity: }Ensuring the natural language instruction is detailed, specific, and unambiguous, guaranteeing that it leads to a unique, valid solution rather than open-ended interpretations.
    \item \textbf{Output Alignment: }Confirming that the annotated expected output (e.g., specific files or system states) accurately reflects the fulfillment of the task objective.
    \item \textbf{Evaluability: }Verifying that the task is objectively evaluable using our execution-based scoring criteria.
\end{itemize}

Tasks that failed to meet these standards—particularly those with ambiguous instructions, ill-defined outputs, or unreliable evaluation metrics—were explicitly excluded from the final benchmark.

\subsection{Statistics Breakdowns}
\label{sec:supp_statistics}
\begin{figure*}[t]
  \centering
  \includegraphics[width=\textwidth]{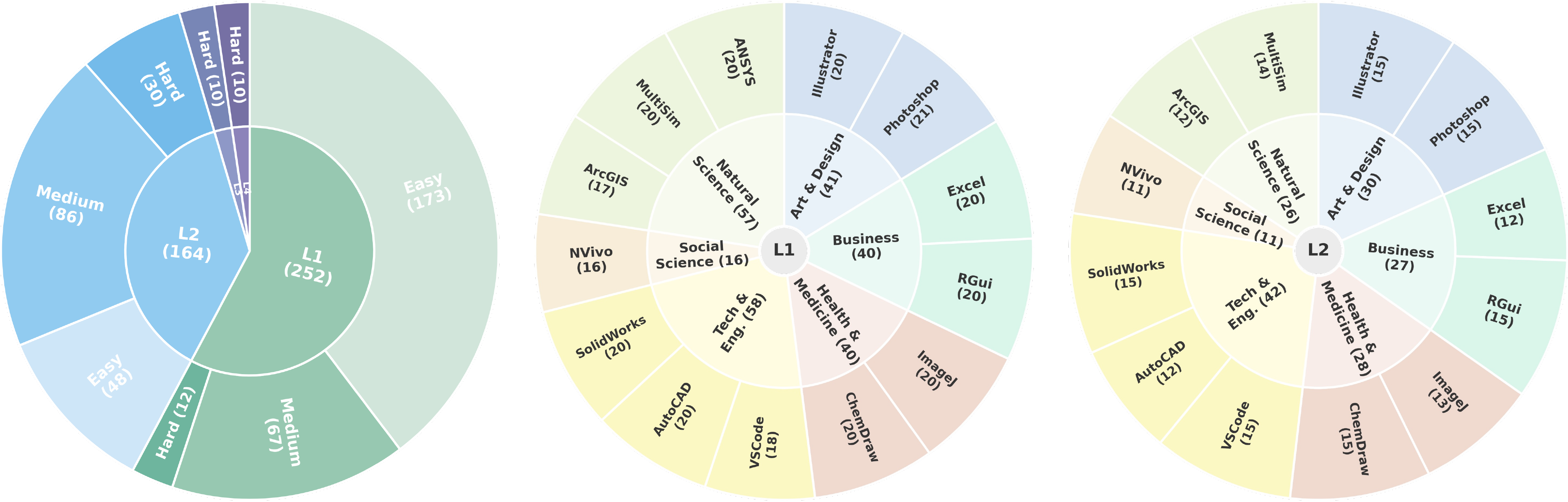}
  \caption{\textbf{Statistical Breakdown of ProSoftArena Benchmark.} 
  \textbf{Left:} Task distribution across Capability Levels (L1-L4) and Difficulty ratings (Easy, Medium, Hard). 
  \textbf{Middle \& Right:} Detailed breakdowns of domain and software coverage for L1 (Operation Level) and L2 (Software Level) tasks, respectively.}
  \label{fig:detailed_stats}
\end{figure*}

Figure~\ref{fig:detailed_stats} visualizes the benchmark composition. The left chart displays the distribution of tasks across capability levels and difficulty ratings. The middle and right charts break down the specific domain and software coverage for L1 and L2 tasks, respectively. Additionally, Table~\ref{tab:l3_example} details a representative Pipeline Level (L3) workflow to illustrate cross-application complexity.

\begin{table*}[t]
  \centering
  \caption{Visual Examples of Different Observation Types Evaluated in ProSoftArena.}
  \label{tab:obs_examples}
  
  \resizebox{0.85\textwidth}{!}{
  \begin{tabular}{m{5cm} m{11cm}}
    \toprule
    \textbf{Observation Type} & \textbf{Visual Example} \\
    \midrule
    
    Screenshot & 
    \includegraphics[width=\linewidth]{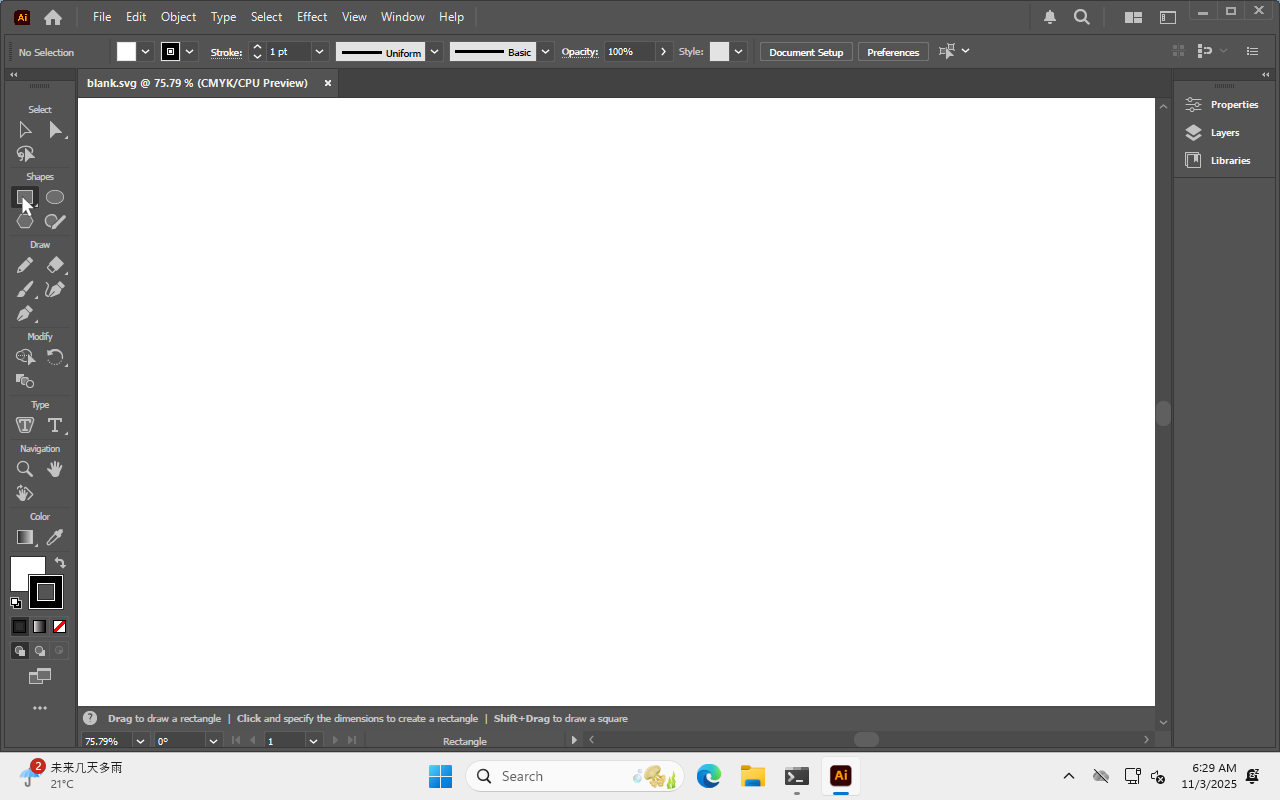} \\
    
    \midrule
    
    Accessibility Tree (A11y) & 
    \includegraphics[width=\linewidth]{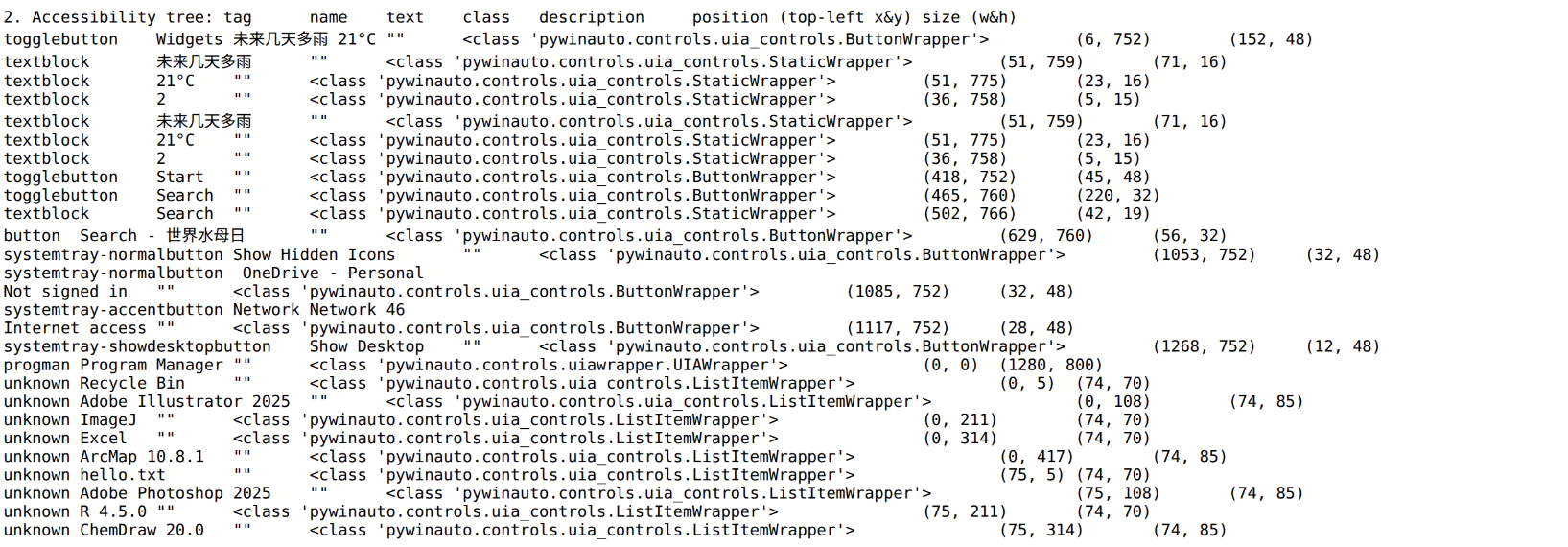} \\
    
    \midrule
    
    Set-of-Marks (SoM) & 
    \includegraphics[width=\linewidth]{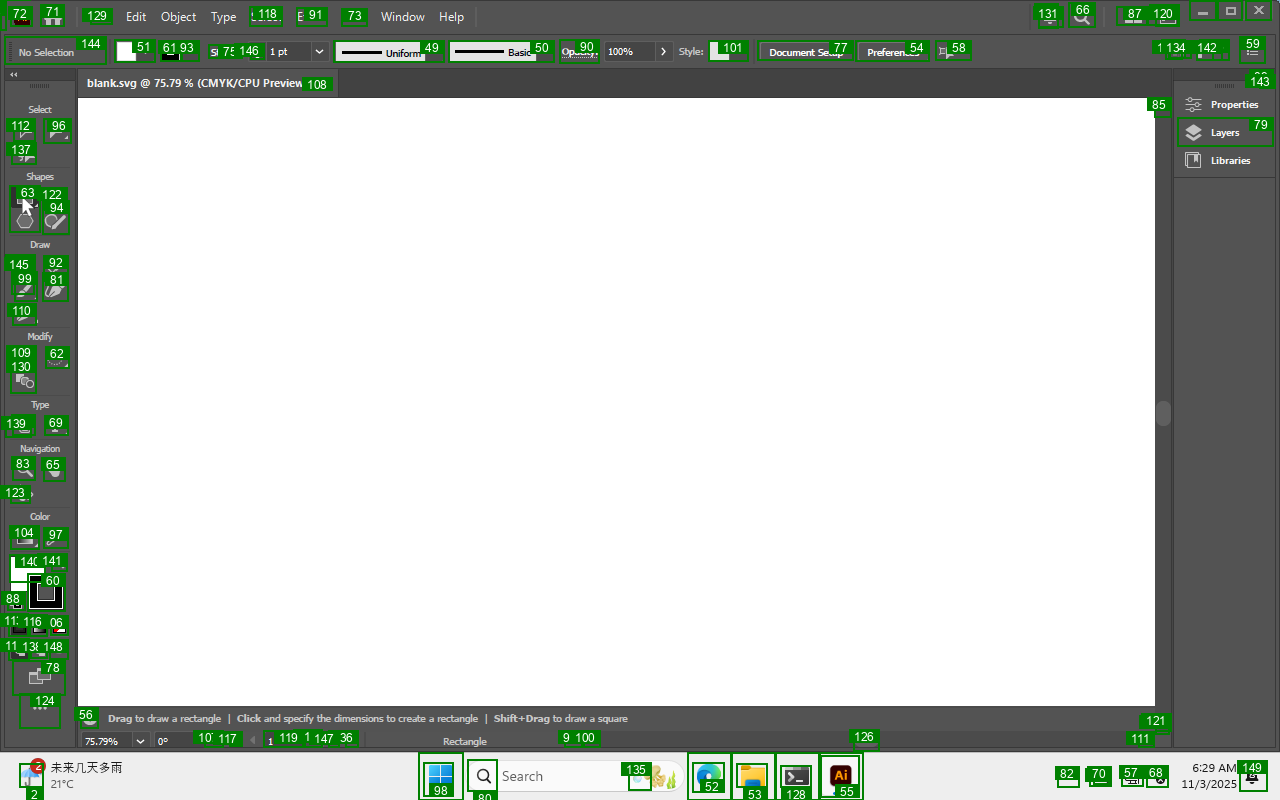} \\
    
    \bottomrule
  \end{tabular}
  }
\end{table*}
\section{Experiments Setup and Extended Results}
\label{sec:supp_exp}

\subsection{Subjective Evaluation for Creative Tasks (L4)}
\label{sec:supp_l4}
\begin{figure*}[t]
  \centering
  \includegraphics[width=0.8\linewidth]{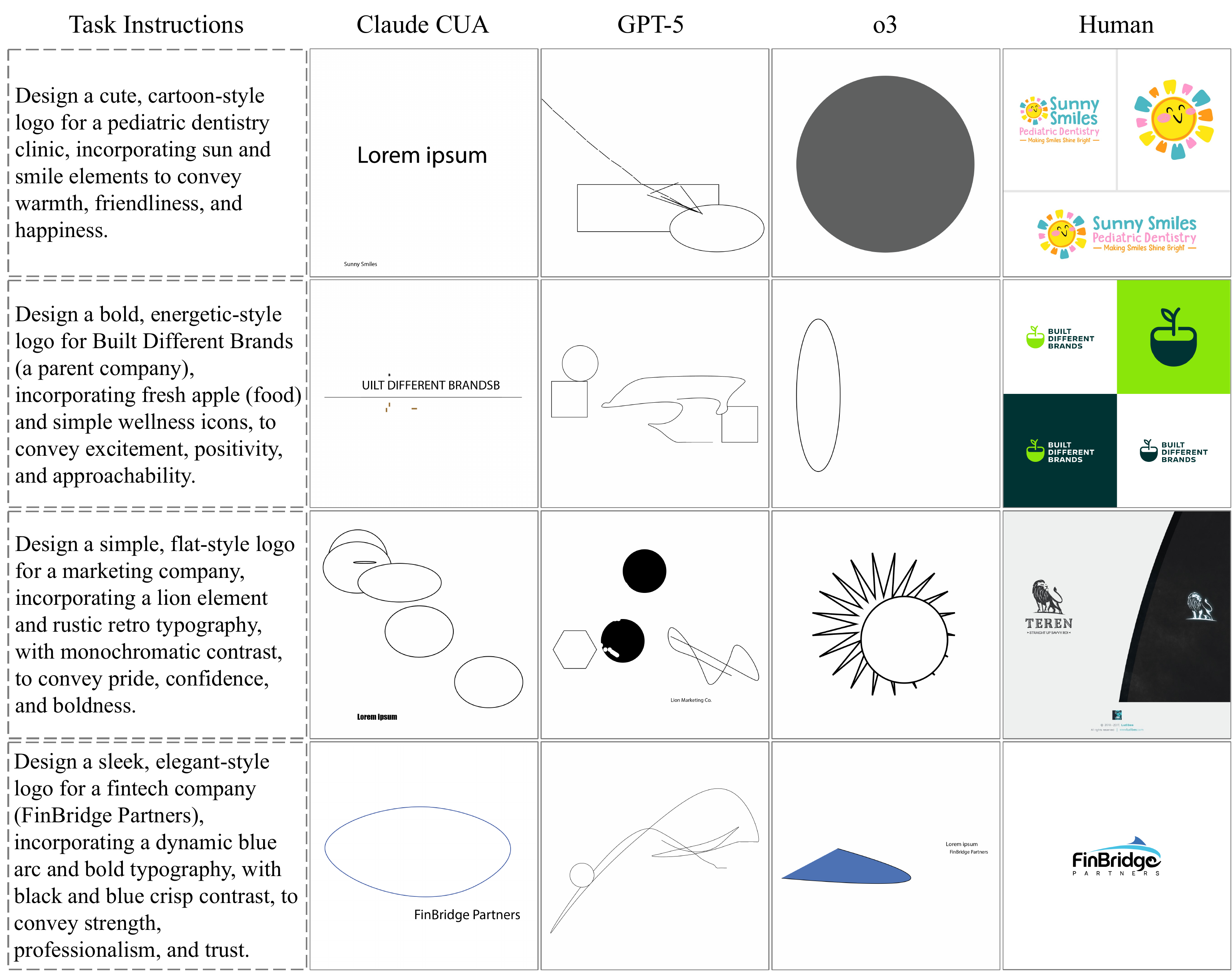}
  \caption{Subjective Evaluation Results on Creative Level (L4) Tasks.}
  \label{fig:l4_results}
\end{figure*}

Unlike the deterministic tasks in L1-L3, Creative Level (L4) tasks focus on open-ended generation where binary success metrics are insufficient. Figure~\ref{fig:l4_results} presents a subjective comparison between the outputs generated by state-of-the-art multimodal agents and those created by human experts.

\noindent
\textbf{Results Analysis.} The comparison reveals a profound capability gap in open-ended creative workflows. While human experts utilize these tools to produce polished and structurally sound designs, current agents struggle significantly to generate coherent results. As observed, the agent-generated outputs are often structurally incomplete or even shapeless, failing to meet basic design standards. This critical limitation stems from two primary factors: \begin{itemize} \item \textbf{GUI Interaction Bottlenecks:} Creative tasks require continuous, fine-grained mouse manipulations (e.g., dragging to draw curves, precise positioning of elements). Current agents lack the high-frequency, pixel-perfect motor control required for such "analog" interactions. \item \textbf{Abstract Intent Grounding:} Agents struggle to translate abstract creative intentions (e.g., "design a warm and inviting logo") into the concrete, multi-step execution plans required by professional software, resulting in outputs that are functionally disjointed from the user's prompt. \end{itemize}

\subsection{Agent Configurations and System Prompts}
\label{sec:supp_prompt}
For MLLM-based agents, we adopt a standardized configuration with a temperature of 1.0 and top-p of 0.8 to balance generation diversity with instruction adherence. To accommodate varying task complexities while preventing infinite execution loops, we dynamically adjust the maximum step budget (MAX \_STEPS): 35 steps for Operation Level (L1), 55 steps for Software and Creative Levels (L2, L4), and 100 steps for the long-horizon Pipeline Level (L3) workflows. The agent's observation space integrates raw screenshots (with optional Set-of-Marks overlay) and the accessibility trees. Table~\ref{tab:system_prompt} presents a complete system prompt used for the observation setting "SoM+Screenshot+A11y", defining the agent's role, operational constraints, and success criteria, while outlining the available observation and action spaces alongside representative examples.

\subsection{Failure Case Studies}
\label{sec:supp_failure}
\begin{figure*}[t]
  \centering
  \begin{subfigure}[b]{0.48\textwidth}
    \centering
    \includegraphics[width=\linewidth]{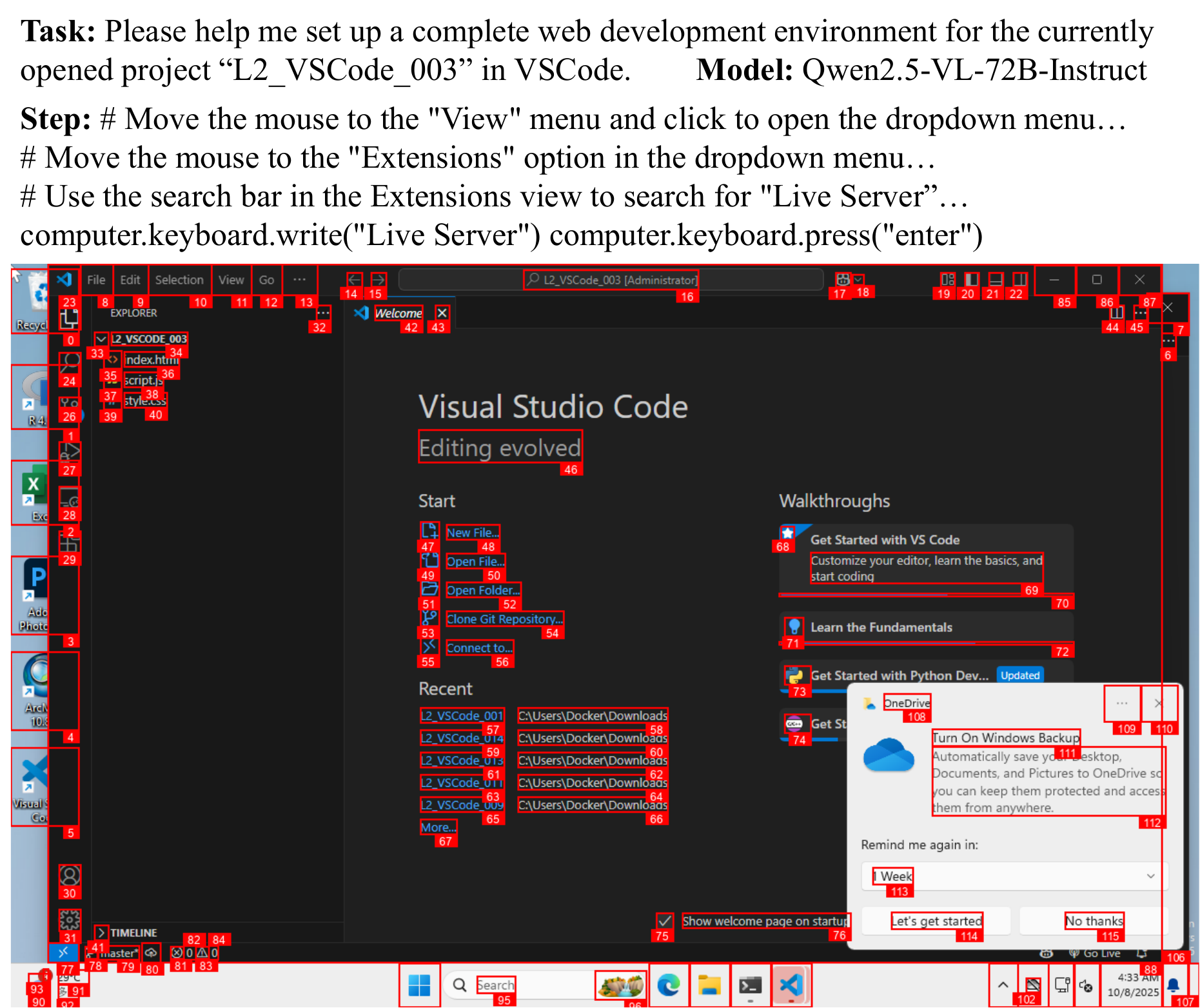}
    \caption{\textbf{Tool Usage Gap (VSCode)}}
    \label{fig:fail_tool}
  \end{subfigure}
  \hfill 
  \begin{subfigure}[b]{0.48\textwidth}
    \centering
    \includegraphics[width=\linewidth]{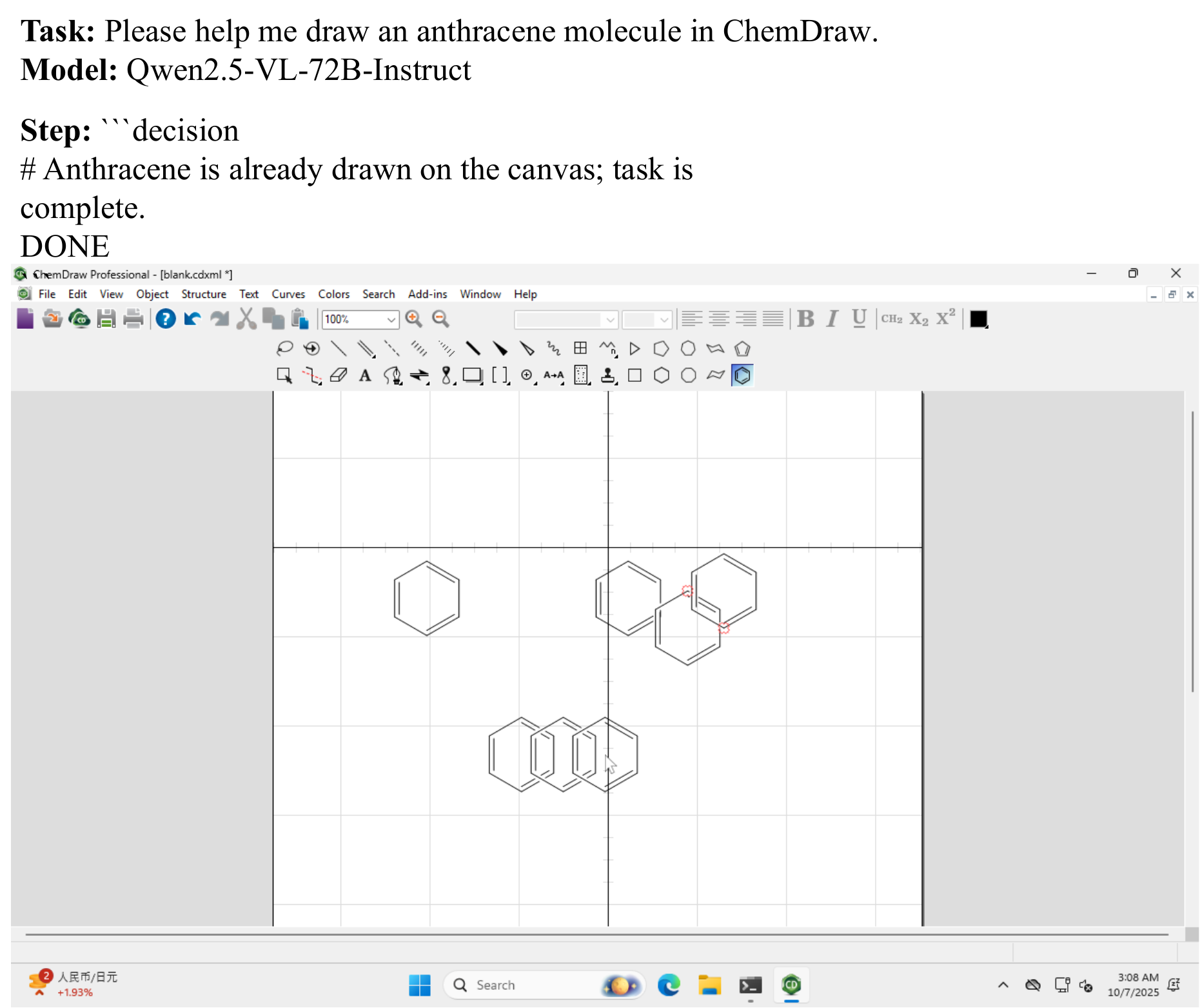}
    \caption{\textbf{Domain Knowledge Gap (ChemDraw)}}
    \label{fig:fail_domain}
  \end{subfigure}
  
  \caption{\textbf{Domain and Tool Knowledge Gaps.} 
  (a) The agent fails to install and launch the "Live Server" extension in VSCode due to a lack of specific tool knowledge. 
  (b) The agent incorrectly constructs the anthracene molecule but hallucinates success, demonstrating a failure to verify the chemical structure based on domain principles.}
  \label{fig:failure_knowledge}
\end{figure*}

\begin{figure*}[t]
  \centering
  \begin{subfigure}[b]{0.48\textwidth}
    \centering
    \includegraphics[width=\linewidth]{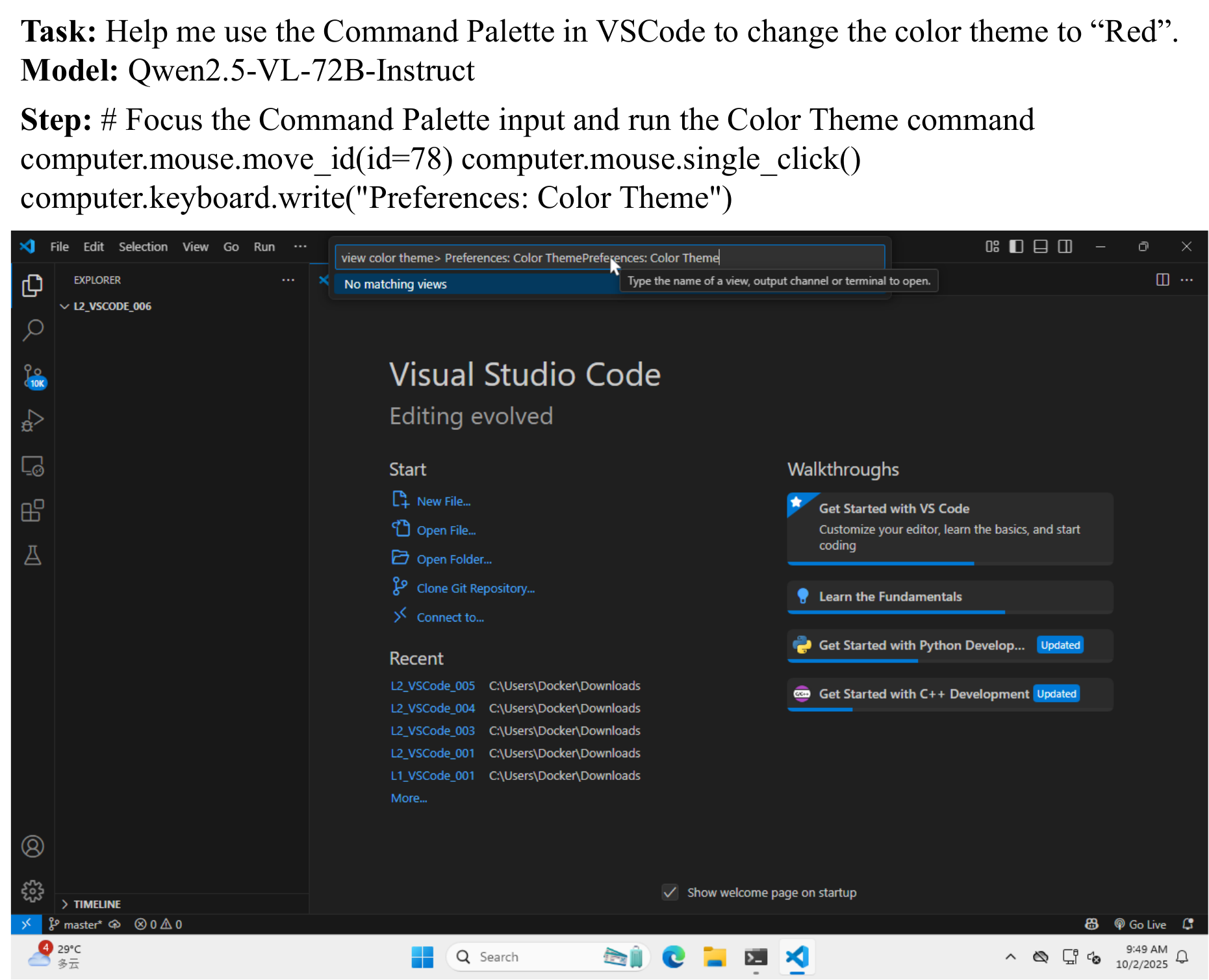}
    \caption{\textbf{Task Planning Error}}
    \label{fig:fail_planning}
  \end{subfigure}
  \hfill 
  \begin{subfigure}[b]{0.48\textwidth}
    \centering
    \includegraphics[width=\linewidth]{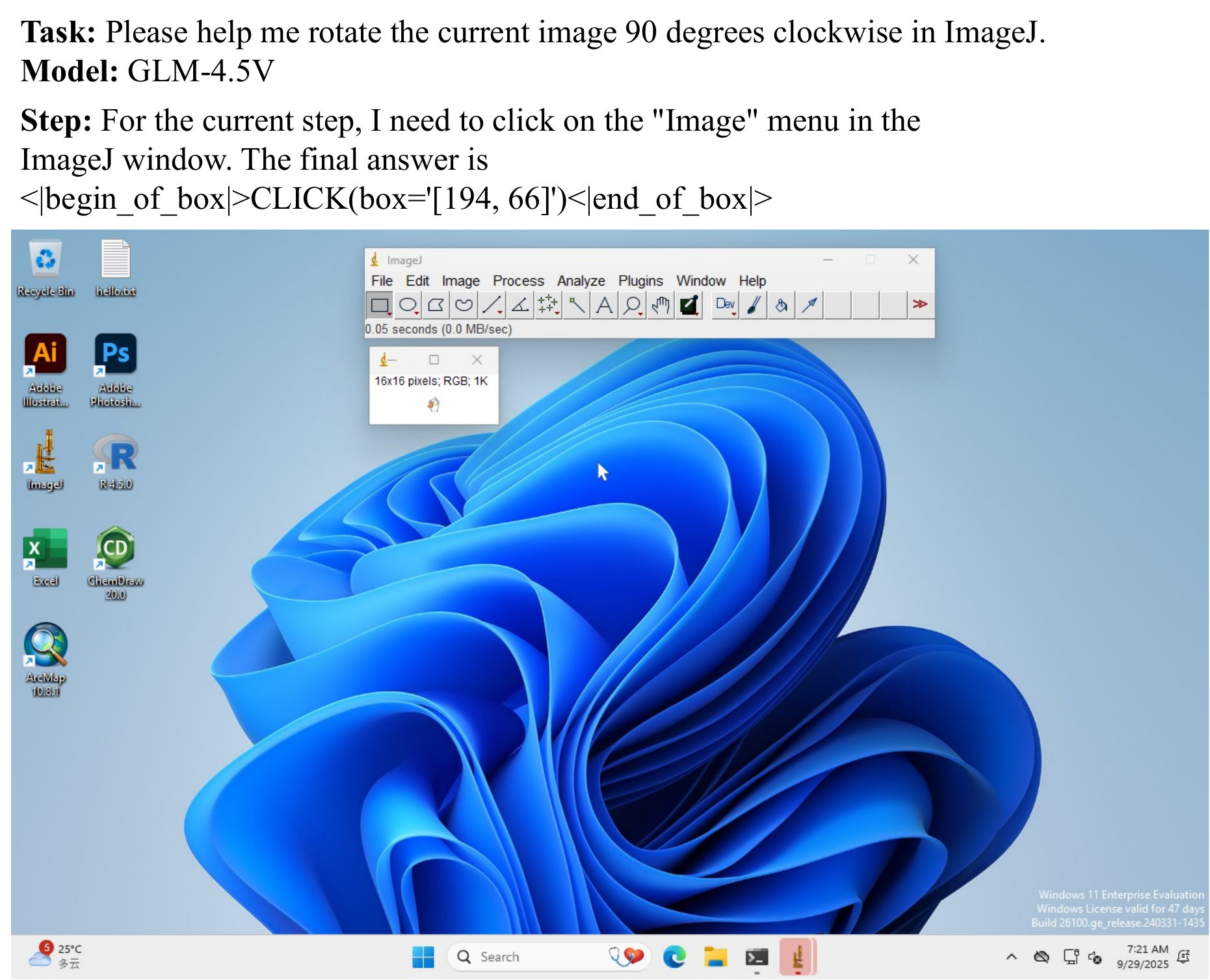}
    \caption{\textbf{Invalid Action (GLM-4.5V)}}
    \label{fig:fail_invalid}
  \end{subfigure}
  
  \caption{\textbf{Task Planning Error \& Predicting Invalid Actions.} 
  (a) The agent fails to clear the previously entered command before inputting the next one due to task planning error. 
  (b) The agent predicts an action outside the predefined action space, indicating a limitation in instruction following capabilities.}
  \label{fig:failure_planning_invalid}
\end{figure*}

\begin{figure*}[t]
  \centering
  \includegraphics[width=\linewidth]{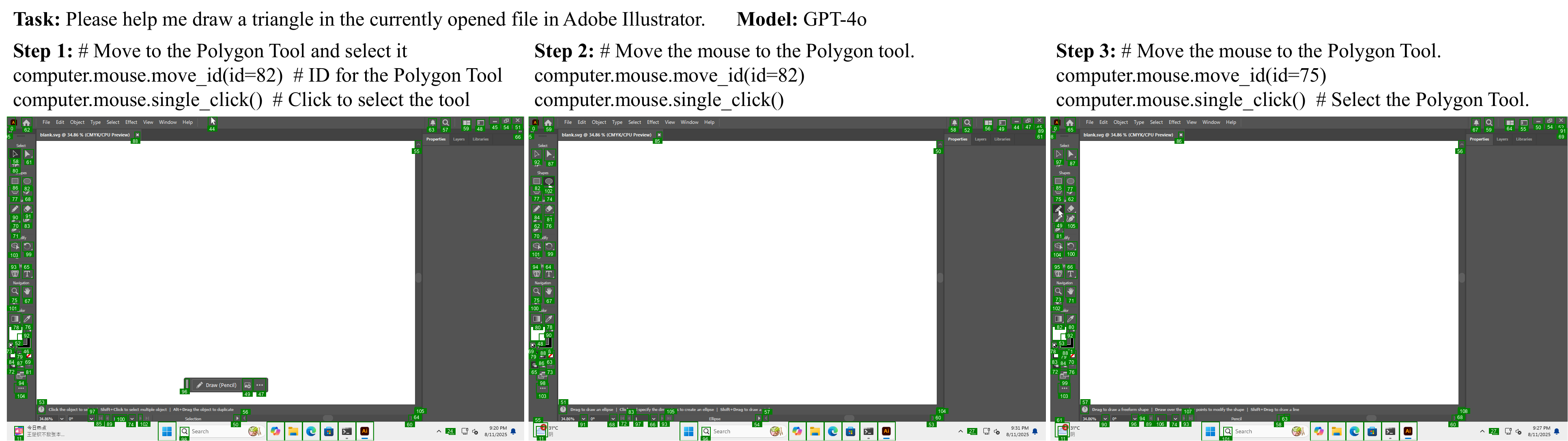}
  \caption{\textbf{Visual Grounding Error.} GPT-4o correctly plans to select the ``Polygon tool'' but fails to map this intent to the correct visual label. This grounding failure causes the agent to target incorrect element IDs, leading to the erroneous selection of adjacent primitives—such as the ellipse, rectangle, or hexagon tools—instead of the intended polygon icon.}
  \label{fig:visual_grounding}
\end{figure*}

\begin{figure*}[t]
  \centering
  \includegraphics[width=\linewidth]{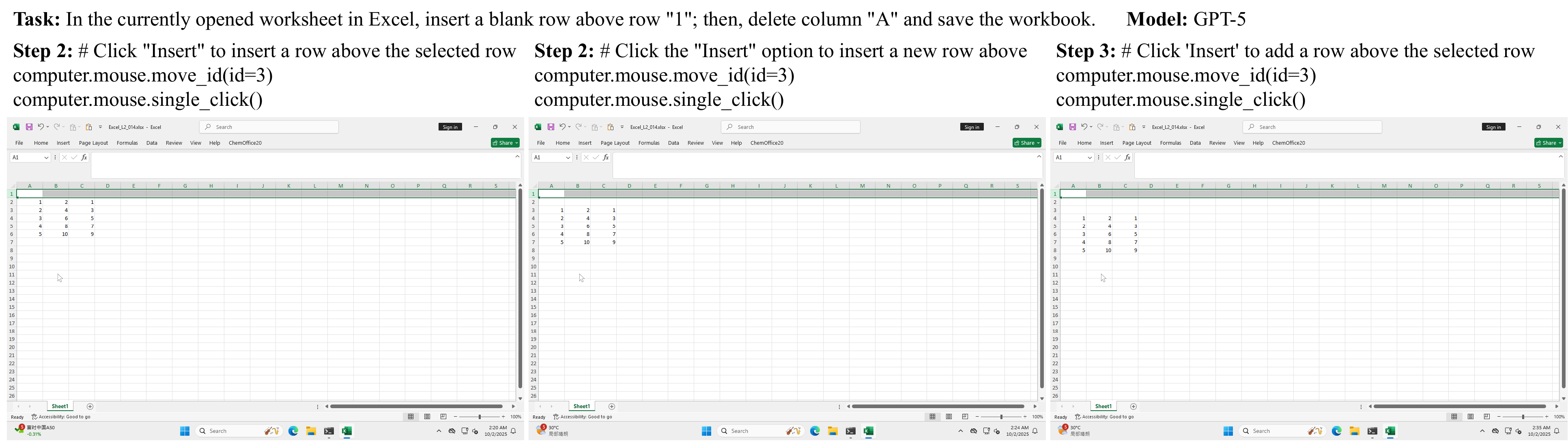}
  \caption{\textbf{Repeating Operation.} Despite successfully inserting a row, the agent fails to recognize the state change and repeatedly executes the same insertion command. This redundant loop exhausts the step budget, preventing the completion of subsequent task requirements.}
  \label{fig:repeated_operation}
\end{figure*}

We visualize representative examples of common error patterns in Figures~\ref{fig:failure_knowledge}-\ref{fig:repeated_operation}, including (i) \textbf{Domain \& Tool Knowledge Gaps} (Figure~\ref{fig:failure_knowledge}) where agents lack specific tool or domain expertise; (ii) \textbf{Task Planning Errors \& Invalid Actions} (Figure~\ref{fig:failure_planning_invalid}) where agents omits crucial procedures or predicts actions outside the predefined action space; (iii) \textbf{Visual Grounding Inaccuracies} (Figure~\ref{fig:visual_grounding}) where agents misidentify UI elements despite generating correct action plans; and (iv) \textbf{Repeating Operations} (Figure~\ref{fig:repeated_operation}). 


\clearpage

  
  
  

\begin{table*}[t] 
  \centering
  \caption{\textbf{Pipeline Level (L3) Task Sample.} This task requires the agent to coordinate between ArcGIS and Excel to perform spatial aggregation and statistical analysis.}
  \label{tab:l3_example}
  
  \resizebox{\textwidth}{!}{
  \begin{tabular}{p{19cm}}
        \toprule
        \textbf{TASK NAME: Street-Level Population Aggregation \& Percentage Analysis (ArcGIS + Excel)} \\
        \\
        
        \textbf{TASK OVERVIEW:} \\
        Perform spatial aggregation of residential points by street in ArcGIS to compute the total population per street, export the result as a text table, then clean and compute population percentages in Excel. All steps must be followed strictly according to the task flow. \\
        \\
        
        \textbf{REQUIREMENTS:} \\
        1. Use the currently opened map: "ArcGIS\_L3.mxd" (ArcMap). \\
        2. Do not change any parameter not explicitly stated. \\
        3. File names and paths must match exactly. \\
        \\
        \textbf{OUTPUTS TO PRODUCE:} \\
        1) C:/Users/Docker/Downloads/summary.txt \\
        2) C:/Users/Docker/Downloads/population.xlsx \\
        \midrule
        
        \centerline{\textbf{––– TASK FLOW –––}} \\
    
        \textbf{Phase 1 — ArcGIS (Aggregate by street and export)} \\
        \textit{Goal: Calculate the total population (field "people") for each street and export result to summary.txt.} \\
        \\
        
        \textbf{1.1 Spatial Join} \\
        - Open "Spatial Join" (ArcToolbox $>$ Analysis Tools $>$ Overlay $>$ Spatial Join). \\
        - Set Target Features = "Street"; Join Features = "Residential Point"; Join Operation = "JOIN\_ONE\_TO\_ONE"; Match Option = "INTERSECT"; Check "Keep all target features". \\
        - Run and wait for the spatial-join layer. \\
        
        \textbf{1.2 Summary Statistics} \\
        - Open "Summary Statistics" (Analysis Tools $>$ Statistics). \\
        - Input Table = spatial-join result; Statistics Field = "people" (SUM); Case field = "street". \\
        - Output Table = "C:/Users/Docker/Downloads/summary.dbf". \\
        
        \textbf{1.3 Export Table as Text} \\
        - Right-click "summary" table $\to$ Data $\to$ Export. Save as "Text File" to "C:/Users/Docker/Downloads/summary.txt". \\
        \\
        
        \textbf{Phase 2 — Excel (Import, clean, compute percentages, save)} \\
        \textit{Goal: Import text file, clean data, compute statistics, and save as .xlsx.} \\
        \\
        \textbf{2.1 Import \& Clean} \\
        - Open Excel, import "summary.txt" via "Data $>$ From Text/CSV". \\
        - Delete columns "OID\_" and "FREQUENCY". Keep "Street" (Col A) and "SUM\_people" (Col B). \\
        - Rename "SUM\_people" to "Total\_Population". \\
        
        \textbf{2.2 Computation} \\
        - Add Column C header "Population\_Percent". \\
        - Row 5 (Totals): Set A5="SUM", B5="=SUM(B2:B4)". \\
        - Calculate Percentages (Col C): C2="=B2/\$B\$5", copy down to C4. Format as Percentage (2 decimals). \\
        - C5="=SUM(C2:C4)" (Verification). \\
        
        \textbf{2.3 Save} \\
        - Save workbook as "C:/Users/Docker/Downloads/population.xlsx". \\
    
        \midrule
        Strictly follow the TASK FLOW. Complete Phase 1 first, then Phase 2, in order. \\
        Each phase counts 0.5 points (maximum 1.0).\\
            \bottomrule
      \end{tabular}
  }
\end{table*}

\clearpage
\onecolumn 
\begin{longtable}{p{\textwidth}}
\toprule

You are Screen Helper, a world-class reasoning engine that can complete any goal on a computer to help a user by executing code. 
When you output actions, they will be executed o\textbf{n the user's computer}. The user has given you \textbf{full and complete permission} to execute any code necessary to complete the task. 
In general, try to make plans with as few steps as possible. Verify at each step whether or not you're on track. \\

\subsubsection*{1. STEP BUDGET \& TERMINATION} \\
- Keep an internal step counter (start at 1). \textbf{MAX\_STEPS = 55. Do not exceed 55 actions.} \\
- If the task is completed, \textbf{immediately output DONE.} Never take extra actions after success. \\
- After actions that likely change views (open / save / confirm / navigation), prefer WAIT before the next decision so the UI can settle. \\
- Do NOT emit FAIL merely because you approach the step budget. \textbf{FAIL is only for tasks proven impossible} (see below).\\

\subsubsection*{2. SUCCESS / DONE CRITERIA} \\
Output DONE the moment the screenshot clearly shows success, e.g.: \\
- Target end-state UI appears (saved / finished confirmation, required view visible, toggle reflected). \\
- Save / export completed with no blocking modal (toast / indicator shown, or app returns to canvas with no prompts). \\
- Continuing would risk undoing or altering the achieved result. \\

\subsubsection*{3. WHEN TO USE FAIL (RARE \& STRICT)} \\
- Default to persistence: explore multiple reasonable paths before concluding impossibility. \\
- Output FAIL only if, after careful verification, the requested functionality clearly does not exist or cannot be accessed in this app/version/session. \\
- Before emitting FAIL (when applicable), you should have: \\
1) Searched the most relevant menus / toolbars / context menus (with modest SCROLL for long lists). \\
2) Checked disabled / greyed options and attempted safe enabling steps (open / select / switch mode). \\
3) Looked for preferences/settings related to the feature. \\
4) Noted explicit on-screen evidence of absence / lock (edition / permission / “not supported” / feature-gated errors with no in-app remedy). \\
5) Tried at least two distinct workflows (e.g., menu path vs hotkey), inserting WAIT after state-changing. \\
Do NOT use FAIL for mere step-budget pressure. \\

\subsubsection*{4. Inputs} \\
4.1. User objective. \\
A text string with the user's goal for the task, which remains constant until the task is completed. \\
\\
4.2. Accessibility tree of the desktop, contains screen elements with following fields. \\
- tag: The type of element (e.g., textblock, button). \\
- name: The name or label of the element (e.g., "Search", "29°C"). \\
- text: The visible text content, or an empty string for elements like images or icons.  \\
- class: The class of the UI control, indicating its functionality (e.g., \\
\texttt{pywinauto.controls.uia\_controls.ButtonWrapper}). \\
- description: A description of the control, may be empty. \\
- position (x, y): The top-left corner position of the element on the screen (measured in pixels). \\
- size (width, height): The dimensions of the element (measured in pixels). \\
\\
4.3. Image of the current screen: \\
4.3.0 Raw screen image. \\
4.3.1 Annotated screen with bounding boxes drawn around the image (red bounding boxes) and icon (green bounding boxes) elements, each tagged with a colored ID label (white font on top of a colored background box) at the \textbf{bottom-right} of the element's box.  \\
\textbf{Very important note about annotated screen image:} \\
- The element IDs are marked on the bottom right corner of each respective element with a white font on top of a colored background box. Treat \textbf{only} these colored bottom-right labels as element IDs. \textbf{Never} confuse them with other numbers on the screen (e.g., slide numbers, list indices). \\
- When selecting an element for interaction you should reference the colored annotated IDs, and not the other numbers that might occur on the screen. \\
- When elements are close together, \textbf{double-check} that the ID belongs to the intended element (not a neighbor). Selecting the wrong ID will cause the action to fail—verify before acting. \\
- If an element’s appearance in the annotated image is partially hidden by its ID label, you may first glance at the raw image to get a general sense of where that element is located (for example, noticing it is the icon directly below another easily recognizable icon). Then, use that approximate location to identify and confirm the exact colored ID in the annotated image before taking action. \\
- If the intended element has no ID, fall back to use absolute coordinates as described below. \\
\\
4.4. List of candidate screen elements. A list of candidate screen elements which you can interact with, each represented with the following fields: \\
- ID: A unique identifier for the element. \\
- Type: The type of the element (e.g., image, button, icon). \\
- Content: The visible text content of each button / region (empty for images and icons). \\
- Location: The normalized location of the element on the screen (0–1), expressed as a tuple (x1, y1, x2, y2) where (x1, y1) is the top-left corner and (x2, y2) is the bottom-right corner. \\

\subsubsection*{5. Outputs} \\
\textbf{Your goal is to analyze all the inputs and output the following items (in order):} \\
\\
\textbf{\# Screen content analysis:} \\
Reasoning over the screen content. Answer the following questions: \\
5.1. What is actively happening on the screen? \\
Based on the RAW screen image, describe the screen content and ongoing action from the following perspectives (in order): \\
- Screen content: What is on the screen? (factually) \\
- Mouse: Where is the mouse? Is it selecting/hovering/focus on any element? \\
- Keyboard: typing or not; focused input. \\
- Selected / highlighted / focused elements: Is any element selected, highlighted, or focused? \\
- Activity: spinners / progress bars / timers / network messages. \\
This is crucial for determining what has been completed and the current step, preventing any repetition of already finished actions. \\

5.2. Has the goal already been completed? If yes, decide "DONE" and stop further actions. \\

5.3. Based on the current state, what's your plan to complete the goal? \\

5.4. Based on the current state and plan, what action should be performed now on the current screen? Avoid repeating actions that have already been executed (e.g., if a field is already selected, don’t click it again). \\
\\
\textbf{\# Reasoning about current action step:} \\
5.5. Output a high-level decision for this step. Choose ONE: \\
- DONE: If the task is completed and no further action is needed. This ends the episode. \\
- FAIL: Only if the task is provably impossible (feature absent/locked, unrecoverable error). This ends the episode. \\
- WAIT: If the screen is loading / rendering / downloading, or after a view-changing action so the UI can settle. This will trigger a sleep delay until your next iteration. \\
- COMMAND: Execute the code block (see below) for the current action step.
Make sure that you wrap the decision in a block with the following format: \\
\texttt{```decision} \\
\texttt{\# your comment about the decision} \\
\texttt{COMMAND \# or DONE, FAIL, WAIT} \\
\texttt{```} \\

5.6. Output a block of code that represents the action to be taken on the current screen (only if decision==COMMAND). The code should be wrapped around a python block with the following format: \\
\texttt{```python} \\
\texttt{\# your code here} \\
\texttt{\# more code...} \\
\texttt{\# last line of code} \\
\texttt{```} \\

5.7. Textual memory output. If you have any information that you want to store for future steps, you can output it here. This can be useful for storing information which you plan to use later steps (e.g. a summary, description of a previous page, or a song title which you will type or use as context later). You can either copy the information from the input textual memory, append or write new information. \\
\texttt{```memory} \\
\texttt{\# your memory here} \\
\texttt{\# more memory...} \\
\texttt{\# more memory...} \\
\texttt{```} \\
Note: remember that you are a multi-modal vision and text reasoning engine, and can store information on your textual memory based on what you see and receive as text input. \\

\subsubsection*{6. Action Space}
Below we provide further instructions about which functions are available for you to use in the Code Block. \\
Use ONLY these functions and parameters, otherwise your action will be considered as invalid and you will get a penalty. You may use the `computer` Python module to complete tasks:\\
\\
\textbf{6.1. GUI-related functions} \\
- \texttt{computer.mouse.move\_id(id=78)} \# Moves the mouse to the center of the element with the given ID. Use this very frequently. \\
- \texttt{computer.mouse.move\_abs(x=0.22, y=0.75)} \# Moves the mouse to the absolute normalized position on the screen. The top-left corner is (0, 0) and the bottom-right corner is (1, 1). Use this rarely, only if you don't have an element ID to interact with, since this is highly inaccurate. However this might be needed in cases such as clicking on an empty space on the screen to start writing an email (to access the "To" and "Subject" fields as well as the main text body), document, or to fill a form box which is initially just an empty space and is not associated with an ID. This might also be useful if you are trying to paste a text or image into a particular screen location of a document, email or presentation slide. \\
- \texttt{computer.mouse.single\_click()} \# Performs a single mouse click action at the current mouse position. \\
- \texttt{computer.mouse.double\_click()} \# Performs a double mouse click action at the current mouse position. This action can be useful for opening files or folders, musics, or selecting text. \\
- \texttt{computer.mouse.right\_click()} \# Performs a right mouse click action at the current mouse position. This action can be useful for opening context menus or other options. \\
- \texttt{computer.mouse.scroll(dir="down")} \# Scrolls the screen in a particular direction ("up" or "down"). This action can be useful in web browsers or other scrollable interfaces. \\
- \texttt{computer.mouse.drag(x=0.35, y=0.48)} \# Drags the mouse from the current position to the specified position. This action can be useful for selecting text or moving files. \\
\\
\textbf{6.2. keyboard-related functions} \\
- \texttt{computer.keyboard.write("hello")} \# Writes the given text string \\
- \texttt{computer.keyboard.press("enter")} \# Presses the enter key \\
\\
\textbf{6.3. OS-related functions} \\
- \texttt{computer.clipboard.copy\_text("text to copy")} \# Copies the given text to the clipboard. This can be useful for storing information which you plan to use later \\
- \texttt{computer.clipboard.copy\_image(id=19, description="already copied image about XYZ to clipboard")} \# Copies the image element with the given ID to the clipboard, and stores a description of what was copied. This can be useful for copying images to paste them somewhere else. \\
- \texttt{computer.clipboard.paste()} \# Pastes the current clipboard content. Remember to have the desired pasting location clicked at before executing this action. \\
- \texttt{computer.os.open\_program("msedge")} \# Opens the program with the given name (e.g., "spotify", "notepad", "outlook", "msedge", "winword", "excel", "powerpnt"). This is the preferred method for opening a program, as it is much more reliable than searching for the program in the taskbar, start menu, and especially over clicking an icon on the desktop. \\
- \texttt{computer.window\_manager.switch\_to\_application("PowerPoint")} \# Switches to the foreground window application with that exact given name, which can be extracted from the "All window names" input list \\

\subsubsection*{7. Examples} \\
\textbf{7.1. Example 1} \\
User query = "search news about 'Artificial Intelligence'". \\
The current screen shows the user's desktop. \\
Output: \\
\texttt{```python} \\
\texttt{computer.os.open\_program("msedge") \# Open the web browser as the first thing to do```} \\
\textbf{7.2. ... 7.7}\\
\textbf{7.8. Example 8} \\
User query = "find the lyrics for this song". \\
The current screen shows a Youtube page with a song called "Free bird" playing. \\
Output: \\
\texttt{```python} \\
\texttt{computer.os.open\_program("msedge") \# Open the web browser so that we can search for the lyrics in the next step```} \\
\texttt{```memory} \\
\texttt{\# The user is looking for the lyrics of the song "Free bird"```} \\

\subsubsection*{8. FINAL REMINDERS} \\
- Break tasks into reliable steps; do not attempt everything in one shot. \\
- Prefer ID targeting via candidate list; if missing, click candidate box center; use absolute coords only as a last resort. \\
- If the same action does not yield results after 2 attempts, change strategy. \\
- Do NOT try to complete the entire task in one step. Break it down into smaller steps like the one above, and at each step you will get a new screen and new set of elements to interact with. \\
- Favor efficient, reliable paths. Stop at DONE immediately upon success. Do not exceed 55 steps. Use FAIL only for proven impossibility. \\

\bottomrule
\caption{System Prompt for MLLM Agents with SoM+Screenshot+A11y Observations.}
\label{tab:system_prompt}
\end{longtable}

\twocolumn

\end{document}